\documentclass[letterpaper,11pt]{article}
\pdfoutput=1

\usepackage{caption}
\usepackage{jheppub}
\usepackage{multirow}
\usepackage{diagbox}
\usepackage{xspace}

\usepackage{subcaption}
\usepackage[countmax]{subfloat}
\usepackage{amssymb}
\usepackage{amsmath}
\usepackage{color}
\usepackage{graphicx}
\usepackage{verbatim}
\usepackage{amsthm}
\usepackage{slashed}
\usepackage{hyperref}
\usepackage{makecell}

\usepackage{side}
\usepackage{ulem}

\allowdisplaybreaks

\newcommand{\MG}{MG5\_aMC@NLO\xspace}

\newcommand{\tmop}[1]{\ensuremath{\operatorname{#1}}}

\preprint{}

\title{
Automated calculation of Jet fragmentation at NLO in QCD
}

\author{ ChongYang Liu$^{1,2}$, Xiao-Min Shen$^{1,2,3}$, Bin Zhou$^{1,2}$, Jun Gao$^{1,2}$}

\affiliation{$^1$INPAC, Shanghai Key Laboratory for Particle Physics and Cosmology, School of
Physics and Astronomy, Shanghai Jiao-Tong University, Shanghai 200240, China}
\affiliation{$^2$Key Laboratory for Particle Astrophysics and Cosmology (MOE), Shanghai 200240, China}
\affiliation{$^3$Deutsches Elektronen-Synchrotron DESY, Notkestr. 85, 22607 Hamburg, Germany}

\abstract{
We present FMNLO, a framework to combine general-purpose Monte Carlo generators and fragmentation functions (FFs). 
It is based on a hybrid scheme of phase-space slicing method and local subtraction method, 
and accurate to next-to-leading order (NLO) in QCD.
The new framework has been interfaced to \MG and made publicly available in this work. 
We demonstrate its unique ability by giving 
theoretical predictions of various fragmentation measurements at the LHC, followed by comparison with the data.
With the help of interpolation techniques, FMNLO allows for fast calculation of fragmentation processes for a large number of different FFs, 
which makes it a promising tool for future fits of FFs.
As an example, we perform
a NLO fit of parton fragmentation functions to unidentified charged hadrons using measurements at the LHC.
We find the ATLAS data from inclusive dijet production show a strong constraining power.
Notable disparities are found between our gluon FF and that of BKK, DSS and NNFF, indicating the necessities of additional constraints and data for gluon fragmentation function.
}
  

\begin{document} 

\maketitle

\section{Introduction}

Fragmentations of quarks and gluons into hadrons have been the central topic of QCD since only hadrons are observed experimentally.
QCD factorization ensures separation of the short and long distance effects into
matrix elements on production of partons and the fragmentation functions(FFs)~\cite{Collins:1989gx, Collins:1998rz, Metz:2016swz}.
In its simplest form, fragmentation functions describe probability distribution on the
fraction of momentum of the initial parton carried by the identified hadron.   
Due to its non-perturbative essential, fragmentation functions 
are usually extracted from fits to a variety of experimental data.
However, dependence of the fragmentation functions on the momentum transfers or the so-called
fragmentation scale follows the Dokshitzer-Gribov-Lipatov-Altarelli-Parisi (DGLAP) evolution equation with time-like splitting kernels. 
For extraction of FFs at next-to-next-to-leading order (NNLO), 
three-loop evolution kernels at ${\cal O}(\alpha_s^3)$ in the strong coupling are needed, 
which have been calculated in Ref.~\cite{Mitov:2006ic,Moch:2007tx,Almasy:2011eq, Chen:2020uvt,Luo:2020epw,Ebert:2020qef}.

The experimental data used to extract FFs includes Single-Inclusive Annihilation (SIA) on lepton colliders, Semi-Inclusive Deep-Inelastic Scattering (SIDIS), and hadron production on hadron-hadron colliders.  
The corresponding parton-level cross sections for SIA have been calculated
at NNLO in Ref.~\cite{Rijken:1996vr,Rijken:1996ns,Mitov:2006wy,Blumlein:2006rr}.
For SIDIS, the next-to-leading order (NLO) corrections are given in
Ref.~\cite{Altarelli:1979kv,Nason:1993xx,Furmanski:1981cw,Graudenz:1994dq,deFlorian:1997zj,deFlorian:2012wk}, 
and approximate NNLO and N$^3$LO corrections have also been obtained from expansion of the resummed expressions\cite{Abele:2021nyo, Abele:2022wuy}.
For $p p$ collisions, the corresponding NLO corrections for single-inclusive production of a hadron are given by Ref.\cite{Aversa:1988vb,deFlorian:2002az,Jager:2002xm,Arleo:2013tya,Kaufmann:2015hma,Chien:2015ctp}.

Various FFs
\cite{bkk1,bkk2,bkk3,
Kniehl:2000fe,
Bourhis:2000gs,Albino:2005me,Hirai:2007cx,
Kretzer:2000yf,
dss1, dss2, dss3, dss4, dss5,
Sato:2016wqj, %\cite{Moffat:2021dji}.
Bertone:2017tyb, Bertone:2018ecm,
Khalek:2021gxf,
Soleymaninia:2022alt} have been extracted from SIA, SIDIS and $p p$ collisions.
However, there exist several limitations in the current tools on calculations of parton
fragmentations at NLO.
First, the available processes of hard scattering are limited and are
usually implemented case-by-case.
Furthermore, interactions in the hard processes are usually constrained to be
SM interactions, 
thus are restrained from applications to study of various new physics beyond the SM (BSM).
Besides, even direct calculations at NLO are too costly in computation time
to be included into a global analysis of fragmentation functions. 
In this work we provide a solution, dubbed FMNLO, by introducing a hybrid scheme of NLO calculations
utilizing a phase-space slicing of collinear regions in combination with the
usual local subtraction methods.
Due to its simplicity we are able to realize the hybrid scheme based on the widely
used program \MG~\cite{Alwall:2014hca,Frederix:2018nkq}.
That ensures numerical calculations on partonic cross sections for arbitrary
hard processes of fragmentations at NLO within SM and BSMs.
We further generate the nominal and convoluted fragmentation functions using
the HOPPET program~\cite{Salam:2008qg,Salam:2008sz} for fast convolutions with the partonic cross sections.
The rest of our paper is organized as follows. 
In Sec.~\ref{sec:framework}, we present the FMNLO 
framework, 
which combines partonic cross section calculation and FFs at NLO QCD,
in a way suitable for Monte Carlo calculation. 
We also show how the corresponding calculation can be boosted 
with interpolation techniques.
In Sec.~\ref{sec:validation}, our framework and its implementation
are validated by comparing our results with
analytic predictions for SIA at lepton colliders
and the predictions of other program for inclusive hadron production at hadron colliders. 
Then we utilize FMNLO to study three cases of hadron production at the LHC
and compare our NLO QCD predictions with the experimental measurements in 
Sec.~\ref{sec:app-LHC}. 
In Sec.~\ref{sec:fit-FFs}, 
more $p p$ collision measurements at the LHC are considered, 
and a NLO fit of parton fragmentation functions to unidentified charged hadron
is performed, 
followed by comparisons of our fitted FFs with existing FFs.
Finally our summary and conclusions are presented in 
Sec.~\ref{sec:conclusions}.

\section{Theoretical framework}
\label{sec:framework}
\subsection{A hybrid scheme}

Cross sections for any infrared and collinear (IRC) safe observables in a
standard subtraction scheme at NLO have the following schematic form:
\begin{align}
	\frac{d\sigma}{dF} & = \int dPS_{m} \Big[|M|_{B, m}^2+|M|_{V, m}^2 + 
	|\tilde{\cal I}|_{m}^2\Big]\delta(\hat F(p_{m};f_{m})-F) \nonumber \\
	& + \int d PS_{m+1} \Big[|M|_{R, m+1}^2 \delta(\hat F(p_{m+1};f_{m+1})-F) - 
	|{\cal I}|_{m+1}^2 \delta(\hat F(\tilde p_{m};\tilde f_{m})-F) \Big],
	\label{eq:dsigmaoverdf}
\end{align}
where $|M|_B^2$, $|M|_V^2$ and $|M|_R^2$ represent square of matrix elements
at leading order (LO), one-loop level and in real corrections, respectively.
$|{\cal I}|_{m+1}^2$ denotes the local subtraction terms constructed 
in $D=4-2\epsilon$ dimensions when using dimensional regularizations, and
\begin{align}
|\tilde{\cal I}|_m^2 = \int PS_1 |{\cal I}|_{m+1}^2
\end{align}
are the integrated subtraction terms over the phase space of real radiations.
We have taken a single differential cross section in observable $F$ as an example
for a process with $m(m+1)$ finale state particles at LO (in real corrections).
One can imagine $F$ being the transverse momentum of either colorless particles
or a clustered jet produced. 
The measure function $\hat F$ for the observable applies on either the Born
kinematics and flavors $\{p_m; f_m\}$, or those in real corrections $\{p_{m+1}; f_{m+1}\}$,
and in real subtractions $\{\tilde{p}_m; \tilde{f}_m\}$.
For local subtraction schemes, for instance, in CS dipole subtraction~\cite{Catani:1996jh,Catani:2002hc} or 
FKS subtraction~\cite{Frixione:1995ms,Frixione:1997np}, contributions from the second line of Eq.~\eqref{eq:dsigmaoverdf}
can be evaluated immediately in four dimensions due to cancellations of both
infrared and collinear singularities.
Note that in the infrared or collinear limits the measure function of
IRC safe observable equals for configurations $\{p_{m+1}; f_{m+1}\}$ and 
$\{\tilde{p}_m; \tilde{f}_m\}$.
Both the virtual corrections and integrated subtraction terms carry poles in
$\epsilon$ which again cancel among each other which renders the first line
of Eq.~\eqref{eq:dsigmaoverdf} being finite in four dimensions.
For fragmentation processes, the complications are due to uncancelled collinear
singularities from splitting of final state partons and thus observables are
not collinear safe.
However, those singularities are universal and can be absorbed into definitions
of bare fragmentation functions similar to the mass factorization in scattering with
initial hadrons.
Considering the transverse momentum distribution of a tagged hadron, a typical
observable in parton fragmentations, one can attempt to use the same formalism as
for IRC safe observable and calculate
\begin{align}
	\frac{d\sigma}{dp_{T,h}} & = \int dx \int dPS_{m} \Big[|M|_{B, m}^2+|M|_{V, m}^2 + 
	|\tilde{\cal I}|_{m}^2\Big]\sum_{i=1}^m\delta(p_{T,h}-x p_{T,i}) D^0_{h/i}(x)\nonumber \\
	& + \int dx \int d PS_{m+1} \Big[|M|_{R, m+1}^2\sum_{i=1}^{m+1}\delta(p_{T,h}-x p_{T,i}) D^0_{h/i}(x)
	\nonumber \\
	&- |{\cal I}|_{m+1}^2 \sum_{\tilde i=1}^m\delta(p_{T,h}
	 -x \tilde{p}_{T,\tilde i}) D^0_{h/\tilde i}(x) \Big].
	 \label{eq:dsigmaoverdpth}
\end{align}
The bare fragmentation functions for finding a tagged hadron ($h$) with a momentum fraction
of $x$ of the mother parton ($i$) can be expressed in terms of the physical ones using
mass factorization with $\overline {\rm MS}$ scheme as
\begin{align}
	D^0_{h/i}(x)=\int^1_{x} {dy\over y} \sum_j P^+_{ji}(y)D_{h/j}(x/y, \mu_D)\equiv \sum_j P^+_{ji}
	\otimes D_{h/j}(x, \mu_D), 
\end{align}
where the time-like convolution kernel can be expressed as
\begin{align}
	P^+_{ji}(y)=\left({4\pi\mu_R^2e^{-\gamma_E}\over \mu_D^2}\right)^{\epsilon}
	\left[\delta_{ij}\delta(1-y)+{\alpha_S(\mu_R)\over 2\pi} P^{+(0)}_{ji}(y){1\over 
	\epsilon}+...\right],
\end{align}
where $\mu_R$ and $\mu_D$ are the renormalization and fragmentation scale respectively, and $\gamma_E$ is the Euler-Mascheroni constant.
The one-loop regularized splitting functions $P^{+(0)}_{ji}$ are given in Appendix~\ref{app:Theory ingredients} and coincide
with those of space-like splitting functions while the two-loop results are available from
Ref.~\cite{Stratmann:1996hn}.
However, one can not take the four dimensional limit in each of the two phase
space integrals and carry out numerical calculations due to the aforementioned
collinear singularities.
Additional subtraction terms and their integrals are needed for the NLO
calculation, for instance, as given in Ref.~\cite{Catani:1996jh,Catani:2002hc}.
In this study we propose a minimal modification of Eq.~\eqref{eq:dsigmaoverdpth} by using additional
slicing of radiation phase space to single out the collinear singularities instead
of including further local subtractions.
We denote that as a hybrid scheme since it involves both methods of local subtractions
and slicing of phase space.
The master formula for the same distribution is given by  
\begin{align}
	\frac{d\sigma}{dp_{T,h}} & = \int dx \int dPS_{m} \Big[|M|_{B, m}^2+|M|_{V, m}^2 + 
	|\tilde{\cal I}|_{m}^2\Big]\sum_{i=1}^m\delta(p_{T,h}-x p_{T,i}) D^0_{h/i}(x)\nonumber \\
	& + \int dx \int d PS_{m+1}(\Theta(\lambda - C)+\Theta(C - \lambda)) 
	\Big[|M|_{R, m+1}^2\sum_{i=1}^{m+1}\delta(p_{T,h}-x p_{T,i}) D^0_{h/i}(x)
	\nonumber \\
	&- |{\cal I}|_{m+1}^2 \sum_{\tilde i=1}^m\delta(p_{T,h}
	 -x \tilde{p}_{T,\tilde i}) D^0_{h/\tilde i}(x) \Big] \nonumber \\
	 & = \int dx \int dPS_{m} \Big[|M|_{B, m}^2+|M|_{V, m}^2 + 
	|\tilde{\cal I}|_{m}^2\Big]\sum_{i=1}^m\delta(p_{T,h}-x p_{T,i}) D^0_{h/i}(x)\nonumber \\
	& + \int dx \int d PS_{m+1}\Theta(C - \lambda) 
	\Big[|M|_{R, m+1}^2\sum_{i=1}^{m+1}\delta(p_{T,h}-x p_{T,i}) D^0_{h/i}(x)
	\nonumber \\
	&- |{\cal I}|_{m+1}^2 \sum_{\tilde i=1}^m\delta(p_{T,h}
	 -x \tilde{p}_{T,\tilde i}) D^0_{h/\tilde i}(x) \Big]+|\tilde{\cal J}|_{m}^2, 
	 \label{eq:master formula}
\end{align}
where we have inserted two $\Theta$ functions to partition into the unresolved and resolved
collinear regions with a cutoff $\lambda$.
The slicing variable $C$ can be chosen as either the minimum of the usual angular
separations of all QCD partons $\Delta \theta=min\{\Delta \theta_{ij}\}$ in the center
of mass frame or the minimum of the boost-invariant angular separation of all
QCD partons $\Delta R=min\{\Delta R_{ij}\equiv \sqrt{\Delta\phi_{ij}^2+\Delta y_{ij}^2}\}$.
The phase space integral of $m+1$-body above the cutoff is free of infrared
and collinear singularities and can be calculated numerically in four dimensions.
The integral below the cutoff can be factorized using the collinear approximations.
The results contain collinear singularities and are included
in terms $|\tilde{\cal J}|_m^2$ which we calculate in the following. 
For NLO calculations, there only exist single unresolved collinear regions.
Precisely, giving the cutoff is small enough, there is no overlap of phase
space between any collinear regions of two partons $\{kl\}$.
We can write the integral below the cutoff as following when neglecting
power corrections that vanish when taking the limit of $\lambda$ to zero,
\begin{align}
	|\tilde{\cal J}|_m^2&=\sum_{\{kl\}}\int dx \int d PS_{m+1}\Theta(\lambda-\Delta R_{kl}) 
	\Big[|M|_{R, m+1}^2\sum_{i=1}^{m+1}\delta(p_{T,h}-x p_{T,i}) D^0_{h/i}(x)
	\nonumber \\
	&- |{\cal I}|_{m+1}^2 \sum_{\tilde i=1}^m\delta(p_{T,h}
	 -x \tilde{p}_{T,\tilde i}) D^0_{h/\tilde i}(x) \Big]
	\nonumber \\
	&= \sum_{i=1}^m \int dx \int d PS_{m}|M|_{B, m}^2 
		\left({\alpha_S(\mu_R)\over 2\pi}\right)\frac{(4\pi\mu_R^2)^{\epsilon}}
		{\Gamma(1-\epsilon)}
	\int_0^1 dz \int_0^{ z (1-z)(\lambda p_{T,i})^2} ds \nonumber \\
&\,\,\,\, 	\frac{[z (1-z) s]^{-\epsilon}}{s}
	\times \sum_j \Big[
		P^{(0)}_{j i}(z,\epsilon) \delta(p_{T,h}-zx p_{T, i})D^0_{h/j}(x) 
	\nonumber \\
	&-{1\over 2} \big(2\delta_{ij}+\delta_{gi}(\delta_{qj}+\delta_{\bar{q}j})\big)
	 P^{(0)}_{j i}(z,\epsilon) \delta(p_{T,h}-x p_{T, i})D^0_{h/ i}(x) 
	\Big].
	\label{eq:Jm}
\end{align}
We have parameterized the phase space of radiations with the invariant mass square $s$ and
the momentum fraction $z$ in each of the collinear regions.
We further use the fact that both the square of real matrix elements and its
subtractions can be written as unregularized splitting functions,
$P^{(0)}_{j i}(z,\epsilon)\equiv P^{(0)}_{ji}(z)+\epsilon P'^{(0)}_{ji}(z)$,
times the square of Born matrix elements.
It is understood that the subscripts except the labels ($q,\,\bar q,\, g$)  represent flavor of that parton ($q,\,\bar q,\, g$) when
appearing in the splitting functions. 
We arrive at a rather compact form for $|\tilde{\cal J}|_m^2$ after carrying out integrals in $s$ and $z$, which is given by
\begin{align}
        |\tilde{\cal J}|_m^2&=\int dx \int d PS_{m}|M|_{B, m}^2 \left({\alpha_S(\mu_R)\over 2\pi}\right)
	\sum_{i=1}^{m}\delta(p_{T,h}-x p_{T,i})\nonumber \\
	&\times \Big[\left({4\pi\mu_R^2 e^{-\gamma_E}\over \mu_D^2}\right)^{\epsilon}
	\left(-{1\over \epsilon }+\ln{\lambda^2 p_{T,i}^2\over \mu_D^2}\right)\sum_j
	P^{+(0)}_{ji}\otimes D_{h/j}(x, \mu_D) + \tilde D_{h/i}(x, \mu_D)\Big],
\end{align}
with $\tilde D_{h/i}(x, \mu_D)\equiv \sum_j I_{ji}\otimes D_{h/j}(x, \mu_D)$, and the
kernel of residuals $I_{ji}(z)$ can be expressed using the unregularized splitting functions,  
\begin{equation}
     I_{ji}(z)= \begin{cases}
	     2\ln [z(1-z)] P^{(0)}_{ji}-P'^{(0)}_{ji} \,, \quad  \{ji\}=qg,\,gq
	     \vspace{0.1in}
    \\
	     \left[2\ln [z(1-z)] P^{(0)}_{ji}-P'^{(0)}_{ji}\right]_+  \,, \quad  \{ji\}=qq
	     \vspace{0.1in}
        \\
	     4C_A \left(\ln [z(1-z)] (z(1-z)+{1-z\over z})+\left({z\ln [z(1-z)]\over 1-z}\right)_+\right)
	     \nonumber \\
	     +{1\over 18}(10C_A+46n_fT_F)\delta(1-z) \,, \quad  \{ji\}=gg
    \end{cases} .
\end{equation}
Substituting $|\tilde{\cal J}|_m^2$ back to the master formula Eq.~\eqref{eq:master formula} one can
find that the remaining collinear divergences or poles in $\epsilon$ cancel with those
from mass factorizations.
After that all remaining pieces are ready for numerical calculations performed directly
in four dimensions as will be explained in the next section.
In above derivations, we have chosen the slicing variable $C=\Delta R$ as an example.
It can be easily transformed into the case of $C=\Delta \theta$ by exchanging
$\lambda p_{T,i}$ with $\lambda E_{i}$ in Eq.~\eqref{eq:Jm}.
We emphasize that the hybrid scheme proposed above can apply equally to processes
without initial state hadrons, e.g., lepton collisions or particle decays, as well
as lepton-hadron or hadron-hadron collisions.
In latter cases, that implies the usual NLO subtraction terms related to initial hadrons
and mass factorizations of parton distributions are included implicitly in the
derivations.
It is interesting to compare our scheme with the two-cutoff method for NLO calculations
of fragmentations introduced in Ref.~\cite{Harris:2001sx}.
We note that our kernels of residuals $I_{gq,qg}(z)$ coincide with similar quantities
therein since the corresponding splittings are free of soft divergences.
For $I_{qq,gg}(z)$, there is no simple correspondence of the two methods since the
soft divergences are handled differently. 

\subsection{Implementation}
The advantage of above scheme is the capability of easy implementations into various
existing programs for NLO calculations designed for IRC safe observables.
We demonstrate that for calculations of differential distribution in the energy
fraction $x_h\equiv 2E_h/Q$ carried by the tagged hadron.
Schematically our master formula can be recast as
\begin{align}\label{eq:remaster}
	\frac{d\sigma}{dx_h} 
	 & = \sum_{i=1}^m \int {dx_i\over x_i} \left[\frac{d\sigma^{(0)}_m}{dx_i}
	 +\frac{d\tilde {\sigma}^{(1)}_m}{dx_i}
	 \right] D_{h/i}(x_h/x_i, \mu_D)
	 %\nonumber \\ &
	  + \sum_{i=1}^{m+1} \int {dx_i\over x_i} 
	  \frac{d\tilde {\sigma}^{(1)}_{m+1}}{dx_i} D_{h/i}(x_h/x_i, \mu_D)\nonumber \\
	 & + \sum_{i=1}^{m} \int {dx_i\over x_i} \left[{\alpha_S(\mu_R)\over 2\pi}
	 \frac{d\sigma^{(0)}_m}{dx_i}
	 \right]
	 \left(\bar D_{h/i}(x_h/x_i, \mu_D)+\tilde{D}_{h/i}(x_h/x_i, \mu_D)\right),
\end{align}
with
\begin{align}
	\bar D_{h/i}(x, \mu_D)\equiv & \left(\ln{\lambda^2 p_{T,i}^2\over 
	  \mu_D^2}\right)P^{+(0)}_{ji}\otimes D_{h/j}(x, \mu_D), 
\end{align}
and $d\sigma^{(0)}_m/dx_i$ is the LO partonic differential cross section with
respect to the energy fraction carried by the $i$-th parton.
The partonic cross sections $d\tilde{\sigma}^{(1)}_m/dx_i$ and $d\tilde{\sigma}^{(1)}_{m+1}/dx_i$ 
can be identified by comparing to various contributions in Eq.~\eqref{eq:master formula}.   
In summary the differential cross sections at NLO and at hadron level can be
expressed as a convolution of the original fragmentation functions ($D$) and its integrals
($\bar D$ and $\tilde D$) with various partonic cross sections.
We have constructed a fast interface specialized for our calculations as following.
First of all the fragmentation function and its integrals at arbitrary scales can be
approximated by an interpolation on a two-dimensional grid of $x$ and $Q$,
\begin{equation}
\label{Eq:grid}
D(x, Q)=\sum_{i=0}^n \sum_{j=0}^n D^{k+i,l+j}I_i^{(n)}\left(\frac{y(x)}{\delta y}-k\right)
	I_j^{(n)}\left(\frac{w(Q)}{\delta w}-l\right),
\end{equation}
where we choose the interpolation variables $y(x)=x^{0.3}$, $w(Q)=\ln(\ln(Q/0.3\,{\rm GeV}))$
and the interpolation order $n=4$.
$D^{k,l}$ is the value on the $k$-th node in $x$ and $l$-th node in $Q$.
The spacing $\delta y$($\delta w$) has been chosen so as to give $N_x=50$ ($N_Q=16$) grid points evenly distributed
for the typical kinematic regions considered.
We use an $n$-th order polynomial interpolating function $I_{i(j)}^{(n)}$ and the
starting grid point $k(l)$ is determined such that $x(Q)$ is located in between the
$k(l)+1$-th and $k(l)+2$-th grid points.
Substituting the interpolated functions to Eq.~\eqref{eq:remaster} we arrive at
\begin{align}
\label{Eq:coe}
	\frac{d\sigma}{dx_h}=\sum_{i=q,\bar q, g}\sum_{k=1}^{N_x}\sum_{l=1}^{N_Q}
	\left(G(x_h)^i_{k,l}D_{h/i}^{k,l}+\bar G(x_h)^i_{k,l}\bar D_{h/i}^{k,l}
	+\tilde G(x_h)^i_{k,l}\tilde D_{h/i}^{k,l} \right).
\end{align}
In practice we calculate partonic cross sections with {\MG}~\cite{Alwall:2014hca,Frederix:2018nkq}
and extract matrix of coefficients $G$, $\bar G$, and $\tilde G$ for a series 
of $x_h$ values. 
The matrices need to be calculated once and stored using histograms in {\MG}.
For arbitrary choices of fragmentation functions we use HOPPET~\cite{Salam:2008qg,Salam:2008sz} to
carry out DGLAP evolution and convolution of fragmentation functions.
Thus the final hadronic cross sections can be obtained via matrix multiplications
efficiently without repeating the calculations of NLO partonic cross sections
which are time consuming.  
Experimental measurements provide bin-averaged cross sections rather than
differential cross sections at a single value of $x_h$.
They can be constructed again using interpolations from differential cross 
sections on a dense grid of $x_h$ which we choose to be the same as the one used for $x$ interpolation of fragmentation functions.
We have verified that the prescribed interpolations on both fragmentation functions and hadronic cross sections give a precision better
than a few per mille in general.
We emphasize that the above fast interface can also work for any
hadronic differential cross sections related to longitudinal momentum
in fragmentations with minimal modifications.
A driver for running {\MG} to generate the NLO coefficient
tables for a variety of distributions and associated fast interface
have been made available as explained in Appendix~\ref{app:install}.

\section{Validation}
\label{sec:validation}
In this section we demonstrate the validity of our calculation scheme and its implementation for several scenarios in both lepton collisions and hadron collisions.
We note that in {\MG} the NLO mode of lepton-hadron collisions is not publicly available yet.
Our calculation scheme can be easily implemented once the
version including lepton-hadron collisions is released.

\subsection{Lepton collisions}

\begin{figure}[htbp]
  \centering
  \includegraphics[width=0.7\textwidth]{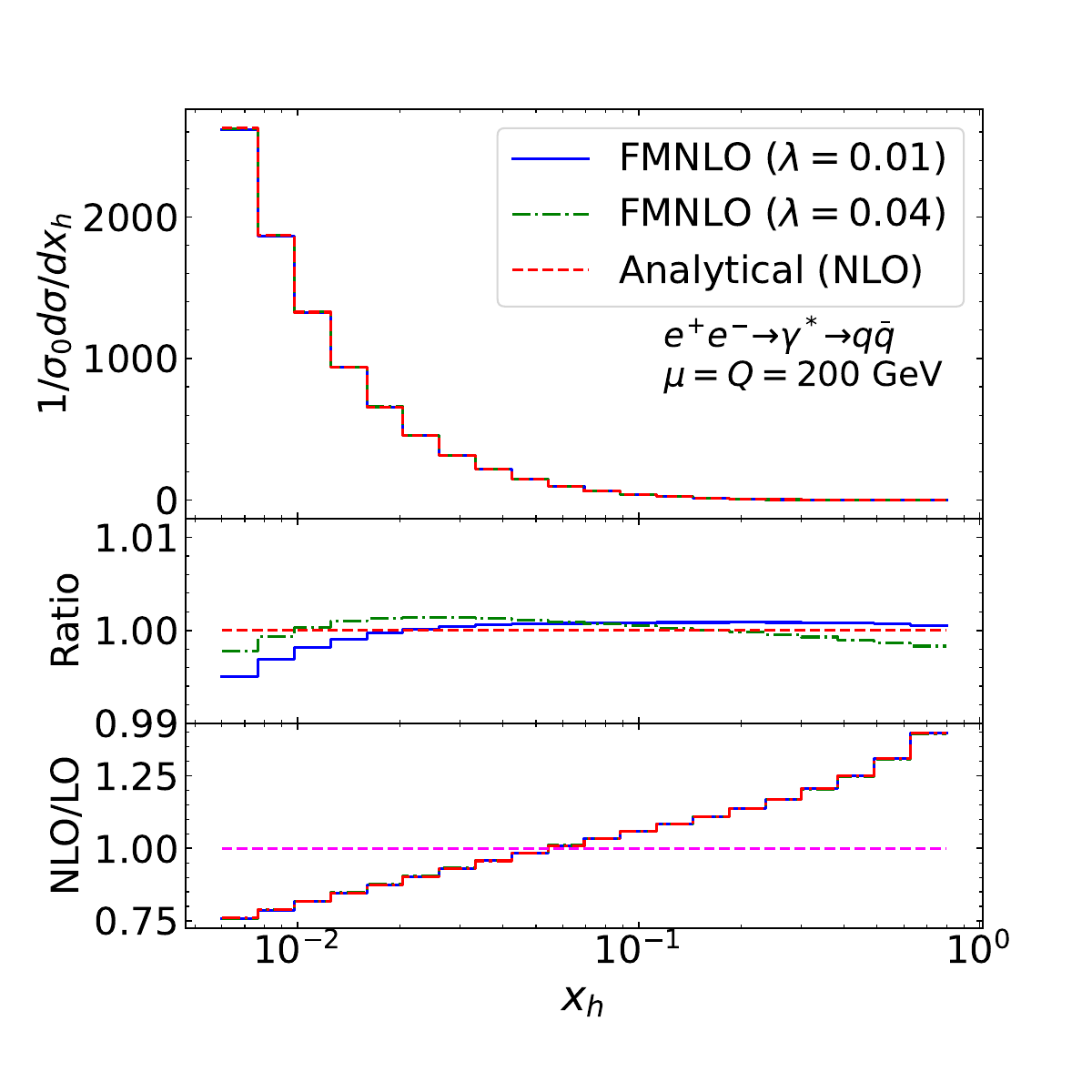}
	\caption{
	Comparison of the NLO predictions on distribution of the hadron energy fraction from FMNLO and from analytical calculations for $e^+ e^- \to \gamma^* \to q \bar{q}$ at a center of mass energy of 200 GeV. 
	}
  \label{Fig:gammaqq}
\end{figure}

\begin{figure}[htbp]
  \centering
  \includegraphics[width=0.7\textwidth]{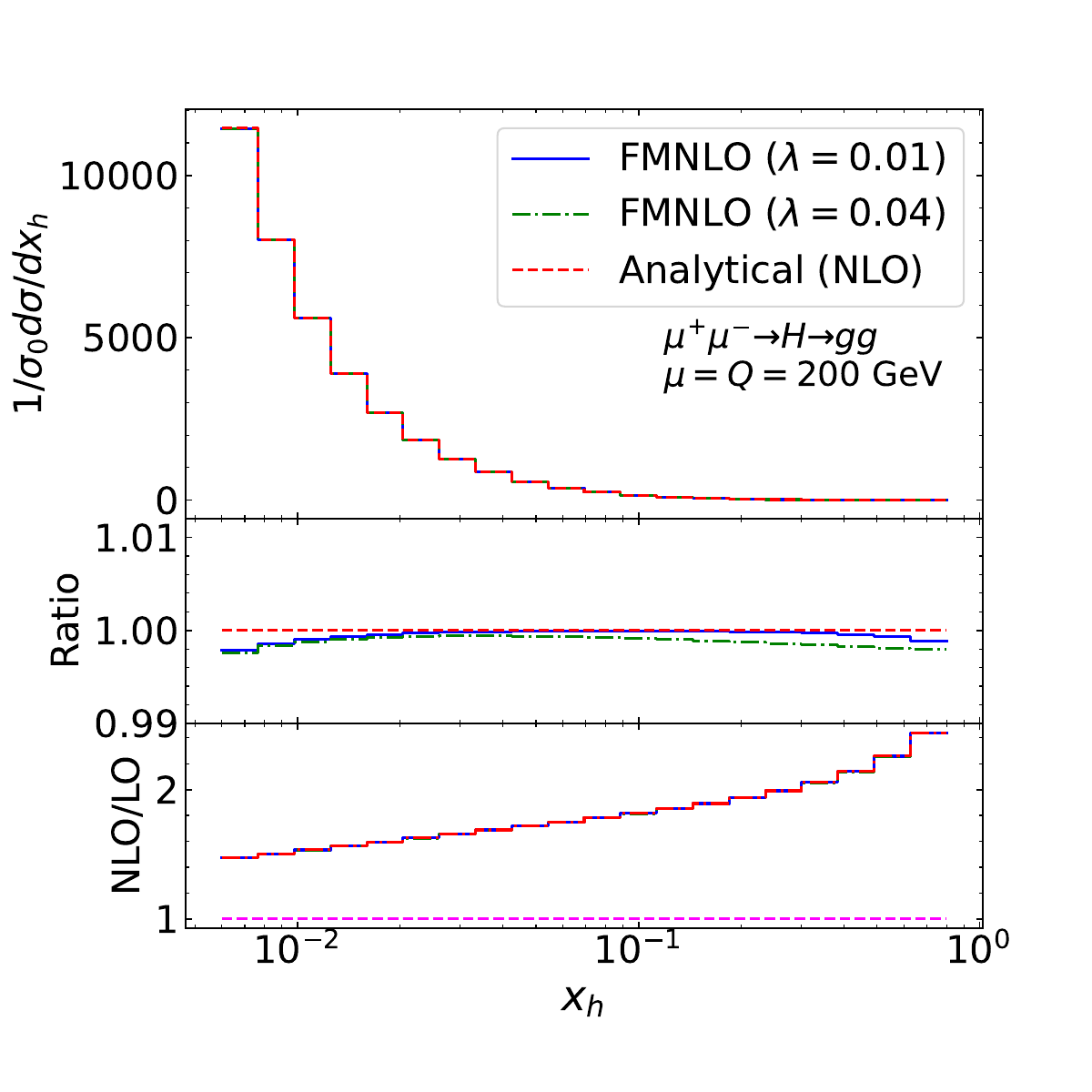}
	\caption{
	Comparison of the NLO predictions on distribution of the hadron energy fraction from FMNLO and from analytical calculations for $\mu^+ \mu^- \to H^* \to g g$ at a center of mass energy of 200 GeV. 
	}
  \label{Fig:hgg}
\end{figure}

We consider two scenarios of lepton collisions for benchmark purpose.
We focus on the NLO predictions for distribution of the energy fraction carried by the tagged hadron, namely $d\sigma/dx_h$.
In the first case, the LO hard process involves the annihilation of an electron-positron pair into quarks through virtual photons. These quarks subsequently undergo fragmentation to produce the tagged hadrons.
In the second case, the LO hard process involves the annihilation of a muon-anti-muon pair into two gluons via the coupling with the SM Higgs boson. These gluons then undergo fragmentation to produce the tagged hadrons.
For above processes the NLO predictions can be calculated analytically with results collected in Appendix~\ref{app:Theory ingredients}.
For simplicity we use a toy model of the fragmentation functions in the
calculations
\begin{equation}
    xD_{h/i}(x,\mu)=N_ix^{-1/2}(1-x)^5,
\end{equation}
with $N_i=1$ and $9/4$ for (anti-)quarks and gluons respectively.
We choose a center of mass energy $Q=200$ GeV and set the renormalization and fragmentation scales to $\mu_R=\mu_D=Q$.
We show comparisons of our numerical results, denoted as FMNLO, and those using NLO analytical formulas in Fig.~\ref{Fig:gammaqq} for the di-quark and in Fig.~\ref{Fig:hgg} for di-gluon production respectively.
We include two groups of results from FMNLO using a cutoff parameter of $\lambda=0.01$ and $0.04$ to check the consistency of our hybrid scheme.
The upper panel shows the NLO predictions on distributions normalized to the LO total cross sections.
The middle and lower panel show the ratios of the three NLO predictions to the analytical results at NLO and the ratios of the NLO results to the LO ones respectively.  
We find very good agreement between our predictions and the analytical results for both channel.
For instance, the NLO predictions with $\lambda=0.04$ differ with the
analytical ones by at most two per mille in the range of $x_h$ from $0.01$ to $1$. 
We have checked that these differences are indeed due to the interpolations used and can be reduced if a denser $x$-grid is used.
The differences between FMNLO predictions with $\lambda=0.01$ and $0.04$
for the virtual photon case are mostly due to fluctuations of Monte Carlo (MC) calculations.
We further compare predictions with a variety of $\lambda$ choice ranging from 0.001 to 0.08 and conclude that the choice of $\lambda=0.04$ is sufficiently small to ensure convergence and stability of MC integration. 
It is worth noting that the numeric effects of the size of a few per mille are much smaller than the typical experimental uncertainties or scale variations of NLO predictions. 

\subsection{Hadron collisions}
We compare our calculations with the INCNLO program~\cite{INCNLO} for the case of unidentified charged hadron production via QCD at hadron collisions. %
We consider the scenario of $pp$ collisions at LHC $7$ TeV and predictions for the transverse momentum distribution of the charged hadrons $d\sigma/ dp_{T,h}$.
The charged hadrons are required to have rapidity $|y|<2.0$.
On various theoretical inputs we use CTEQ6M NLO parton distributions~\cite{Pumplin:2002vw} and the BKK NLO fragmentation functions~\cite{bkk1,bkk2,bkk3}.
Furthermore, for simplicity, we fix both the renormalization and factorization scales to $100$ GeV, and set the fragmentation scale to
$p_{T,h}$, namely transverse momentum of the charged hadron.

\begin{figure}[htbp]
  \centering
  \includegraphics[width=0.7\textwidth]{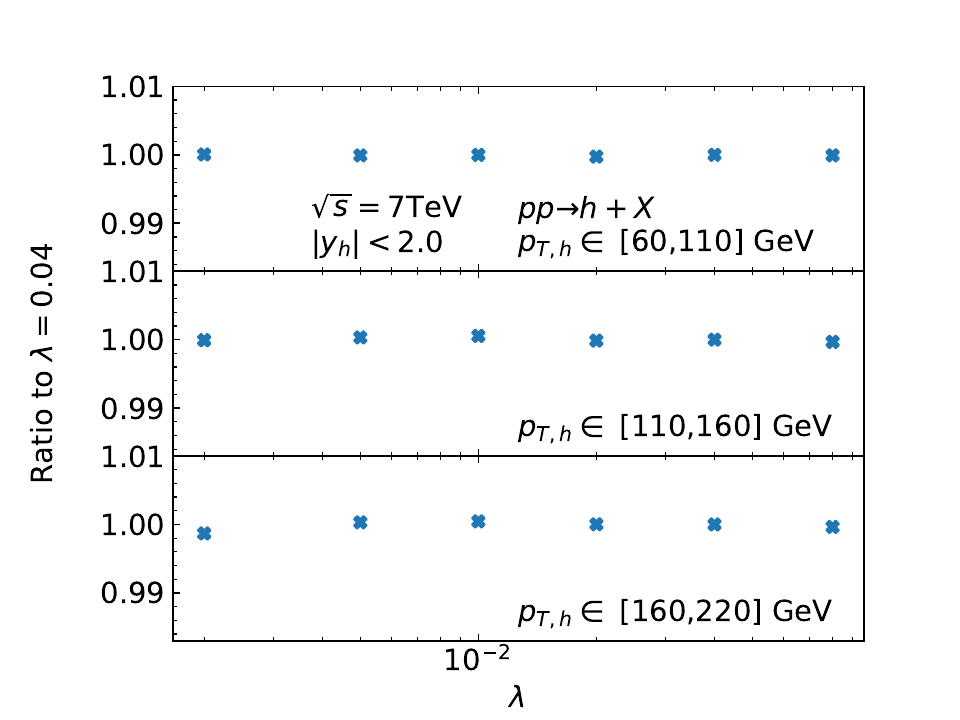}
	\caption{
	Ratio of the NLO predictions on distribution of the hadron transverse momentum from FMNLO with different choices of cut-off parameter relative to $\lambda=0.04$  for inclusive jet production in $pp$ collisions with a center of mass energy of 7 TeV.
	Three representative bins on the transverse momentum have been selected, and a rapidity cut of $|y_h|<2$ of hadrons has been applied.
	}
  \label{Fig:lambda-dep}
\end{figure}

\begin{figure}[htbp]
  \centering
  \includegraphics[width=0.7\textwidth]{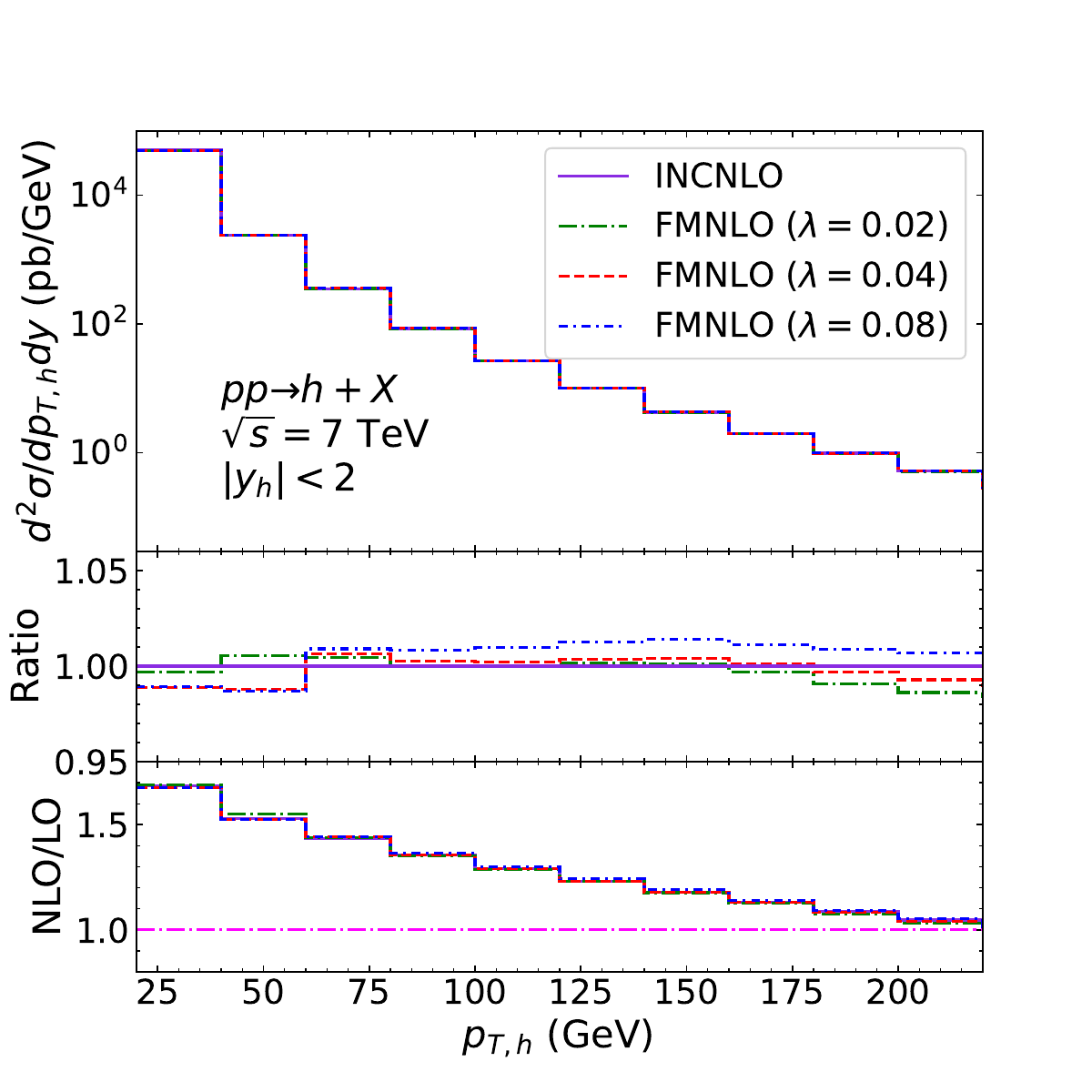}
	\caption{
	Comparison of the NLO predictions on distribution of the hadron transverse momentum from FMNLO and from INCNLO for inclusive jet production in $pp$ collisions with a center of mass energy of 7 TeV.
	A rapidity cut of $|y_h|<2$ of hadrons has been applied.
	}
  \label{Fig:incnlo-fmnlo}
\end{figure}

In Fig.~\ref{Fig:lambda-dep} we demonstrate independence of our NLO predictions on the choice of the cutoff parameter.
We show predictions of the double differential cross sections for $p_{T,h}$ in three kinematic bins from $60$ to $220$ GeV, for several choices of $\lambda$ from $0.002$ to $0.08$.
The variations are within $1\%$ in general mostly due to uncertainties of MC integration. 
In practice, we recommend using a value of $0.02\sim 0.04$ for numerical stability. 
In Fig.~\ref{Fig:incnlo-fmnlo} we present comparisons of our NLO predictions with those from INCNLO1.4~\cite{INCNLO} for a finer binning.
The agreements of the three predictions with $\lambda=0.02$, $0.04$, and $0.08$ are similar to that in Fig.~\ref{Fig:hgg}.
From the middle panel we can find our predictions with $\lambda=0.04$ agree with INCNLO predictions at a few per mille in general. 
In the lower panel ratios of the NLO to LO predictions for
various conditions is presented.
The NLO corrections can reach to 70\% in lower $p_{T,h}$ regions which are much larger than the discrepancies mentioned. 

\section{Applications at the LHC}
\label{sec:app-LHC}

Prescribed calculation scheme and its numerical implementation are especially desirable for predictions of various measurements carried out at the LHC.
In typical fragmentation measurements at the LHC, the requirement is often imposed that the tagged hadron is produced either within a reconstructed jet or in association with an isolated photon or a $Z$ boson.
Meanwhile, jet algorithms and various selection cuts are applied in the analyses which can be implemented easily in a MC event generator such as \MG. 
In the following, we show three examples of such calculations that we adapted to the corresponding LHC measurements. 
We focus on the measurements of spectrum of unidentified charged hadrons.
Predictions on tagged hadrons with a specified flavor can be obtained easily using the same NLO grids multiplied with the corresponding fragmentation functions.    

In the following calculations we use the CT14 NLO parton distribution functions~\cite{Dulat_2016}, the BKK fragmentation functions~\cite{bkk1,bkk2,bkk3} and the NNFF1.1 fragmentation functions~\cite{Bertone:2017tyb, Bertone:2018ecm} for unidentified charged hadrons.
We set central values of the factorization and renormalization scales ($\mu_{F,0}$ and $\mu_{R,0}$) to the default dynamic scale used in \MG, namely the sum of the transverse mass of all final state particles divided by 2.
For the fragmentation scale, we set its central value ($\mu_{D,0}$) to the maximum of the transverse momentum of all final state particles.  
The above central values equal in the case of only two massless particles in the final states.
The scale variations are obtained by taking the envelope of theory predictions of the 9 scale combinations of $\mu_F/\mu_{F,0}=\mu_R/\mu_{R,0}=\{1/2,1,2\}$ and $\mu_D/\mu_{D,0}=\{1/2,1,2\}$.
We note alternative choices on the fragmentation scale of using the transverse momentum of the jet multiplied by the jet cone size when calculating hadron fragmentation inside the jet~\cite{Kaufmann:2015hma}. %
For typical jet cone sizes of $\sim 0.5$ used in the LHC measurements, the choice is close to our nominal choice of the fragmentation scale. 

\subsection{Isolated-photon-tagged jets} \label{sec:photon}

\begin{figure}[htbp]
  \centering
  \includegraphics[width=0.7\textwidth]{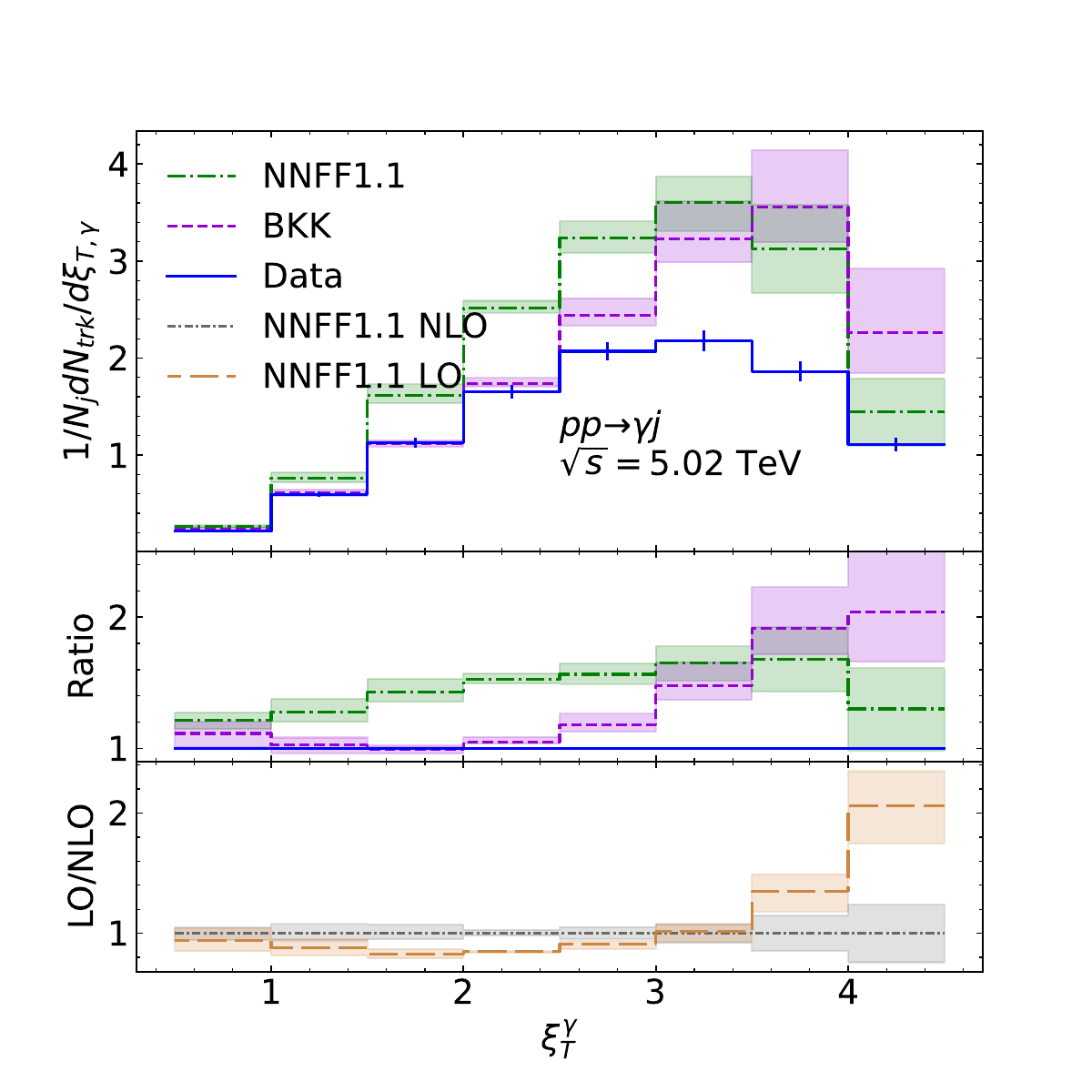}
	\caption{
	Comparison of NLO predictions to CMS measurement on normalized distribution of $\xi_T^{\gamma}$ for isolated photon production in $pp$ collisions with a center of mass energy of 5.02 TeV.
	The two colored bands represent predictions including scale variations, based on NNFF1.1 and BKK fragmentation functions respectively.
	The error bars indicate the total experimental uncertainties.
	Theoretical predictions have been normalized to the central value of data in the middle panel.
	In the lower panel the two bands correspond to LO and NLO predictions based on NNFF1.1, normalized to the NLO prediction with nominal scale choice.
	}
  \label{Fig:1801-04895}
\end{figure}

In Ref.~\cite{1801.04895} the CMS collaboration measured parton fragmentation based on hard scattering events in $pp$ collisions ($\sqrt s=5.02$ TeV) consisting of an isolated photon in association with jets. 
The photon is required to have a transverse momentum $p_{T,\gamma}>$ 60 GeV and a pseudo-rapidity $|\eta_{\gamma}|<$ 1.44.
Jets are clustered with anti-$k_T$ algorithm~\cite{Cacciari:2008gp} with $R=0.3$ and are required to have $p_{T,j}>$ 30 GeV and $|\eta_{j}|<$ 1.6.
They select jets that have an azimuthal separation to the photon $\Delta \phi_{j\gamma}>7\pi/8$ and analyze the charged-particle tracks inside the jet with transverse momentum $\vec p_{T,h}$ in Ref.~\cite{1801.04895}.
The charged tracks are required to have $p_{T,h}>1$ GeV.  
The transverse momentum of the photon $\vec p_{T,\gamma}$ serves as a good reference of the initial transverse momentum of the fragmented parton.
Thus $\xi_T^{\gamma} \equiv \ln [-p_{T,\gamma}^2/(\vec p_{T,\gamma}\cdot \vec p_{T,h})]$ is a good probe of the momentum fraction carried by the
charged hadron. 
The results are presented in a form of $1/N_{j}dN_{trk}/d\xi_T^{\gamma}$
which is simply a linear combination of the quark and gluon fragmentation functions at the LO evaluated at a momentum fraction of $e^{-\xi_T^{\gamma}}$.
Fig.~\ref{Fig:1801-04895} comprises three panels that present a comparison between NLO predictions obtained from different fragmentation function sets and experimental data. 
The first panel displays the results derived from the BKK and NNFF1.1 sets. 
Upon examination, it becomes evident that the results from the BKK set closely resemble the experimental data and exhibit a good agreement in the lower $\xi_T^\gamma$ region, ranging from $0.5$ to $2.5$. 
As $\xi_T^\gamma$ increases, the discrepancy enlarges, which can be attributed to the lack of fitted data in the small $x=p_{T,h}/p_{T,\gamma}$ ($<0.01$) regions~\cite{bkk1,bkk2,bkk3}.
This observation is further supported by the second panel, where the results are normalized to the experimental data. 
The NNFF1.1 results match the experimental data in the lower and higher regions, while a significant deviation is observed in the middle region.
Moreover, the error band indicates that the theoretical uncertainties increase with higher values of $\xi_T^\gamma$.
Finally, in the last panel, 
we give the LO and NLO predictions based on NNFF1.1, normalized to NNFF1.1 results at NLO with nominal scale choice.
%the impact of NLO calculations on the LO prediction is illustrated.
%
The ratios indicate that the NLO corrections contribute insignificantly, less than 20\%, in the region $\xi_T^\gamma<3.5$. 
However, for $\xi_T^\gamma>3.5$, the contribution switches from negative to positive, and the ratio rises rapidly to nearly 2.

\subsection{$Z$ boson tagged jets}

We now turn to the relevant calculations for the production process of Z boson in association with jets at LHC.
In Ref.~\cite{2103.04377} the CMS collaboration measured parton fragmentation based on hard scattering events of the above process, where $\sqrt s=5.02$ TeV.
The $Z$ boson is required to have a transverse momentum $p_{T,Z}>$ 30 GeV, and no jet reconstructions are performed. 
They also analyzed all charged-particle tracks with an azimuthal separation to the $Z$ boson $\Delta \phi_{trk,Z}>7\pi/8$ in Ref.~\cite{2103.04377}.
The charged tracks are required to have $p_{T,h}>1$ GeV and $|\eta_h|<$ 2.4.  
Different from the production process of an isolated photon in association with
jets, we use a distribution of $1/N_{Z}dN_{trk}/dp_{T,h}$ to show our results.

\begin{figure}[htbp]
  \centering
  \includegraphics[width=0.7\textwidth]{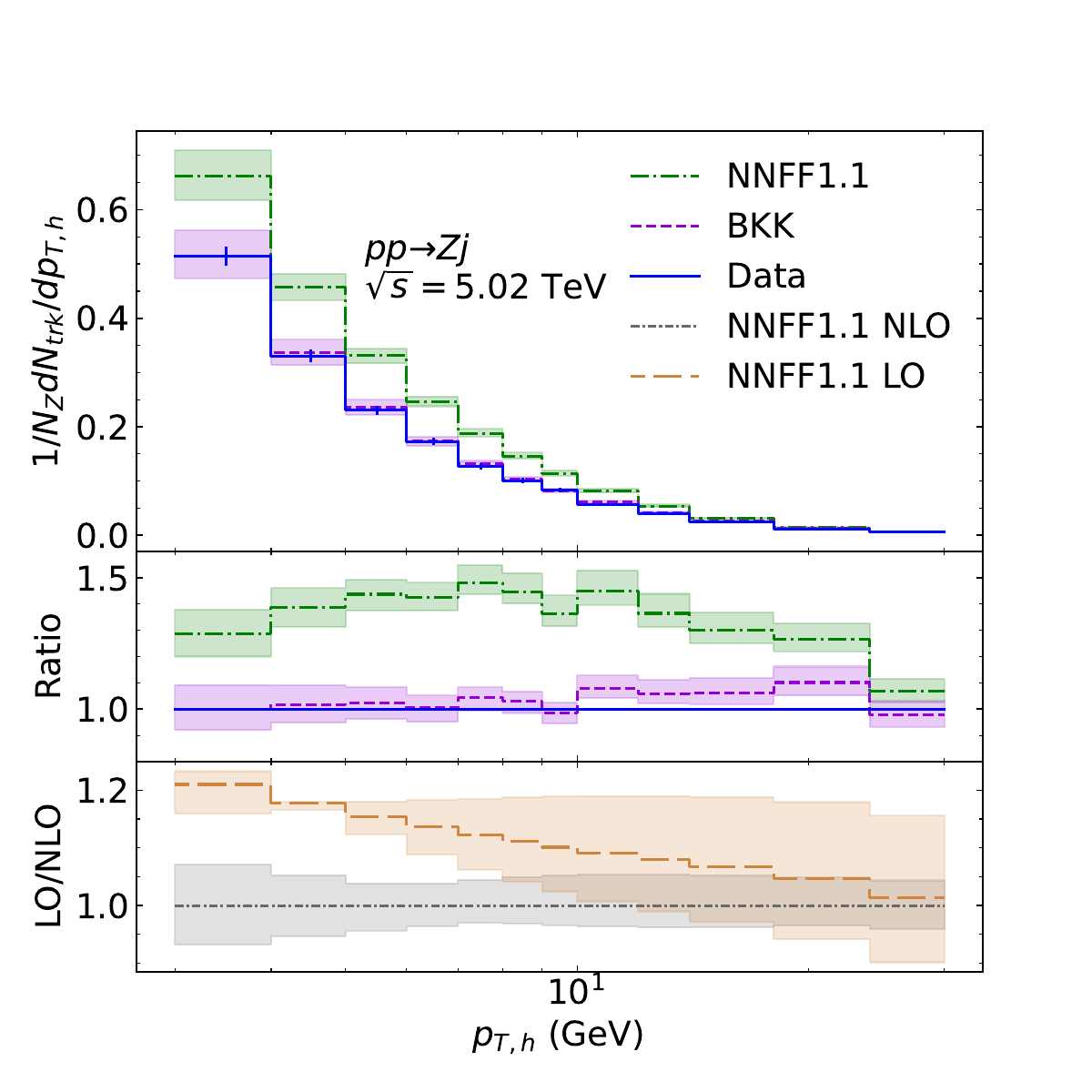}
	\caption{
	Similar to Fig.~\ref{Fig:1801-04895} but with CMS measurement on normalized distribution of $p_{T,h}$ for $Z$ boson production in $pp$ collisions with a center of mass energy of 5.02 TeV.  
	}
  \label{Fig:2103-04377}
\end{figure}

In Fig.~\ref{Fig:2103-04377}, the BKK and NNFF1.1 results are depicted as previously mentioned.
It is apparent from the first panel that the BKK data exhibits better agreement with the experimental data in the whole kinematic region.
In the second panel, we find that, in most regions, the experimental data lies within the error band of the BKK results, with a maximum deviation of approximately 20\%.
Meanwhile, the NNFF1.1 results show a greater discrepancy, particularly in the middle region.
In the third panel, it can be seen that, in most regions, the NLO corrections are negative, but they diminish as $p_T$ increases.
The maximum corrections at NLO is approximately 20\%.

\subsection{QCD inclusive dijets}

In this subsection, we present the third example of the calculations mentioned above.
In Ref.~\cite{1906.09254} the ATLAS collaboration measured parton fragmentation based on hard scattering events in $pp$ collisions ($\sqrt s=13$ TeV) consisting of two or more jets.
Jets are clustered with anti-$k_T$ algorithm with $R=0.4$ and are required to have $p_{T,j}>$ 60 GeV and $|\eta_{j}|<$ 2.1.
The two leading jets are required to satisfy a balance condition $p_{T,j1}/p_{T,j2}<$ 1.5, where $p_{T,j1(2)}$ are the transverse momentum of the (sub-)leading jet.
They also analyzed charged-particle tracks inside the jet classified according to its transverse momentum and pseudo-rapidity (forward or central) in Ref.~\cite{1906.09254}.
The charged tracks are required to have $p_{T,h}>0.5$ GeV and $|\eta_h|<$ 2.5.
The results are presented in a differential cross section of $1/N_{j}dN_{trk}/d\zeta$ with $\zeta\equiv p_{T,h}/p_{T,j}$ and $p_{T,j}$ being the transverse momentum of the jet probed\footnote{We note that the distributions presented in the experimental publication have been multiplied by the bin width of each data points.}.

\begin{figure}[htbp]
  \centering
  \includegraphics[width=0.7\textwidth]{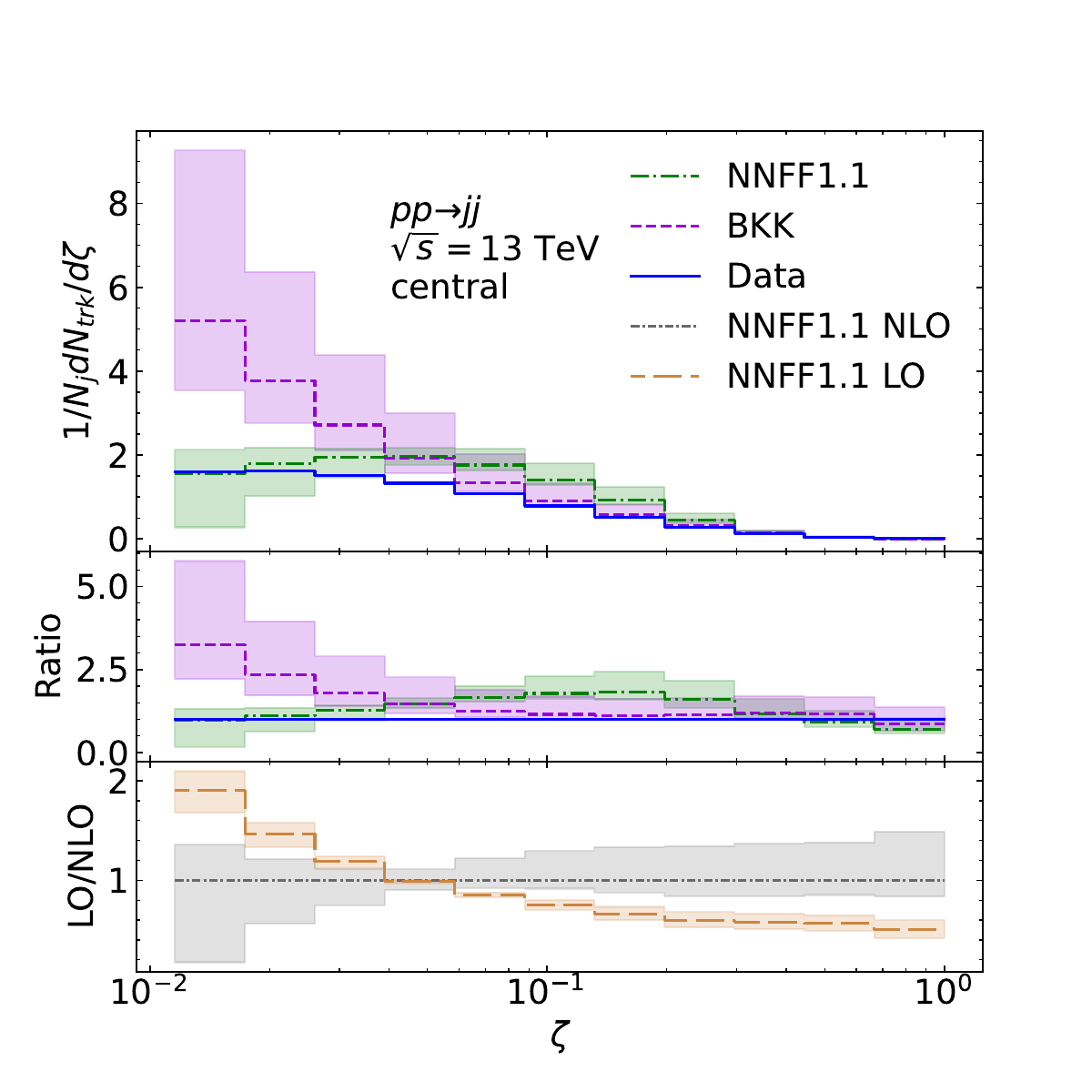}
	\caption{
	Similar to Fig.~\ref{Fig:1801-04895} but with ATLAS measurement on normalized distribution of $\zeta$ for dijet production in $pp$ collisions with a center of mass energy of 13 TeV.  
	}
  \label{Fig:1906.09254}
\end{figure}

We present our NLO predictions and compare them to the ATLAS measurement using the central jet of the two leading jets and with $p_{T,j}\in[200,\,300]$ GeV in Fig.~\ref{Fig:1906.09254}.
The data are displayed as mentioned before.
From the first two panels,
we find both the NNFF1.1 and BKK results fit well in the high $\zeta$ region. However, the BKK data aligns more closely with the experimental data.
In the lower $\zeta$ region, it can be seen that the first three bins of the NNFF1.1 data exhibit a closer resemblance. 
And the error band of the BKK results in these regions is considerable. 
In the third panel, it is apparent that the NLO correction is more significant compared to the previous two experiments, and the ratio can reach nearly 2. The NLO correction transitions from negative in the lower region of $\zeta$ to positive in the higher region.

\section{Analysis of fragmentation functions}
\label{sec:fit-FFs}

In this section we perform a NLO fit of the parton fragmentation functions to unidentified charged hadrons using a variety of experimental data from $pp$ collisions at the LHC.
Those include processes on production of charged hadrons from inclusive dijets, in association with an isolated photon and in association with a $Z$ boson.
They can probe fragmentation of both gluon and quarks in a wide
kinematic region due to different production mechanisms involved.
We demonstrate that such a fit at NLO accuracy with a few hundreds of experimental data points can be accomplished easily with the help of the FMNLO framework.
In the following we first briefly introduce our selection of experimental data sets and the fitting framework, and then show our best-fit and the estimated uncertainties of the fragmentation functions.

\subsection{Experimental data sets}

In this study, we analyzed several recent publications on fragmentation function measurements at the LHC over the past five years.
Relevant information including the kinematic coverage are summarized in Table.~\ref{Tab:data_summary}. 
We focus our analysis solely on data obtained from $pp$ collisions.
These studies were conducted at a center of mass energy of 5.02 TeV, with the exception of the ATLAS inclusive dijet analysis which used a higher energy of 13 TeV. 
The measurements can be separated into three categories including using an isolated photon or a $Z$ boson recoiling against the fragmented parton, or using the clustered jet as a reference of the fragmented parton.

\begin{table}[]
  \centering
\begin{tabular}{|c|c|c|c|c|}
\hline
\textbf{Experiments} & \textbf{lum.} & \textbf{observables}                                      & \textbf{N$_{pt}$} & \textbf{Range} \\ \hline
CMS 5.02 TeV                                 & 27.4 $\mathrm {pb}^{-1}$                                        & $1/N_j dN_{trk}/d\xi_{T}^{\gamma}$ \cite{1801.04895}      & 8(5)           & $\xi_{T}^{\gamma}\in $[0.5,\,4.5]        \\ \hline
ATLAS 5.02 TeV                                & 25 $\mathrm {pb}^{-1}$                                          & $1/N_j dN_{trk}/dp_{T,h}$ \cite{1902.10007}  & 10(7)          & $p_{T,h}\in $[1,\,100] GeV  \\ \hline
CMS 5.02 TeV                                 & 320 $\mathrm {pb}^{-1}$                                          & $1/N_Z dN_{trk}/dp_{T,h}$ \cite{2103.04377}     & 14(11)          & $p_{T,h}\in $[1,\,30] GeV  \\ \hline
ATLAS 5.02 TeV                                  & 160 $\mathrm {pb}^{-1}$                                          & $1/N_Z d^2N_{trk}/dp_{T,h}d\Delta\phi$ \cite{2008.09811}  & 15(9)           & $p_{T,h}\in $[1,\,60] GeV  \\ \hline
ATLAS 13 TeV                                  & 33 $\mathrm {fb}^{-1}$                                         & $1/N_j dN_{trk}/d\zeta$ (central) \cite{1906.09254}  & 261(143)           & $\zeta\in $[0.002,\,0.67]   \\ \hline
ATLAS 13 TeV                                    & 33 $\mathrm {fb}^{-1}$                                         & $1/N_j dN_{trk}/d\zeta$ (forward)\cite{1906.09254}     & 261(143)           & $\zeta\in $[0.002,\,0.67]   \\ \hline
\end{tabular}
	\caption{
	Summary on experimental data sets used in this analysis, including the observable measured, the number of data points before and after data selection, and the kinematic range covered. 
	}
  \label{Tab:data_summary}
\end{table}

In the case of the isolated-photon-tagged jets, the CMS 2018 analysis~\cite{1801.04895} measured the normalized distribution $1/N_{j}dN_{trk}/d\xi_T^{\gamma}$ that has been explained in previous sections.
They probe a region of momentum fractions of the parton carried by hadrons from $0.01\sim 0.6$ based on the definition of $\xi_T^{\gamma}$.
The ATLAS 2019 analysis~\cite{1902.10007} has different setups as the CMS analysis.
We highlighted a few of them as below.
Firstly the photons and the jet are required to have a transverse momentum in [80, 126] GeV and [63, 144] GeV respectively.
The pseudo-rapidity of the photons and the jets have been extended to 2.37 and 2.1 compared to the CMS measurement.
In addition they measured the normalized distribution $1/N_{j}dN_{trk}/dp_{T,h}$ in the region $p_{T,h} \in [1, 100]$ GeV, that is used in our fit.
%from 1 to 100 GeV, that is used in our fit. 
%

%
For measurements involving $Z$-tagged jets, the CMS 2021 analysis~\cite{2103.04377} measured the normalized distribution $1/N_{Z}dN_{trk}/dp_{T,h}$ with setups detailed in previous sections.
In the ATLAS 2020 analysis~\cite{2008.09811}, the same distribution was measured for three transverse momentum regions of the $Z$ boson, namely [15, 30] GeV, [30, 60] GeV, and beyond 60 GeV.
Besides, the requirement on azimuthal separation between the charged track and the $Z$ boson is $\Delta \phi_{trk,Z}>3\pi/4$ instead. 
The covered $p_{T,h}$ region is [1, 30] GeV and [1, 60] GeV for CMS and ATLAS respectively.
Lastly in the ATLAS 2019 analysis of inclusive dijets at high energy and with high luminosity \cite{1906.09254}, they measured the normalized distribution $1/N_{j}dN_{trk}/d\zeta$ detailed in previous sections.
That covered momentum fractions of the parton carried by hadrons from 0.002 to 0.67, as well as a wide range of the transverse momentum of the partons by utilizing jets in finned bins of $p_{T,j}$ from 100 to 2500 GeV.
Furthermore, the distributions are measured independently for the central and forward jet of the two leading jets to increase further the discrimination on fragmentation of gluon and quarks.  
Despite of the wide coverage on momentum fraction or transverse momentum of the hadrons from above measurements, on the theoretical predictions it requires a careful evaluation on validity of the factorization framework and on stability of the perturbation expansions. 
There have been previous studies~\cite{d_Enterria_2014,thennpdfcollaboration2019charged} showing difficulties on fitting to experimental data in certain kinematic regions indicating large higher-order corrections or even violations of collinear factorization.
In this study we take a conservative approach by selecting only those data points corresponding to momentum fractions $x>0.01$ at LO and data points with transverse momenta of the hadrons $p_{T,h}>4$ GeV.
Furthermore, we exclude the jet transverse momentum region of [100, 200] GeV for the inclusive dijet measurements since that corresponds to a low transverse momentum of the hadrons in general.
Similarly, we exclude the $\xi_{T}^{\gamma}$ regions greater than 3 for the CMS isolated-photon-tagged measurement.
These kinematic selections reduce our total number of data points from 569 to 318 as can be seen from Table.~\ref{Tab:data_summary}.
In principle one can perform a scan on above kinematic selections and study the stability of the fitted fragmentation functions which we leave for future investigation.

\subsection{Framework of the fit}
The parameterization form of fragmentation functions to unidentified charged hadrons used at the initial scale $Q_0$ is
\begin{equation}\label{eq:para}
xD_{h/i}\left(x, Q_{0}\right)=a_{i,0} x^{{\alpha}_i}(1-x)^{\beta_i} \left(1+\sum_{n=1}^p a_{i,n}x^n\right),
\end{equation}
where $\{\alpha, \beta, a_{n}\}$ are free parameters in the fit.
We choose $Q_0=5$ GeV and use a zero-mass scheme for heavy quarks with $n_f=5$.
We assume fragmentation functions equal for all quarks and anti-quarks
since the data sets we selected show weak sensitivity on quark flavors of the fragmented partons.
The degree of polynomials is set to $p=2$ since improvements of fit by introducing higher-order terms are marginal.
Thus the total number of free parameters is 10.
The fragmentation functions are evolved to higher scales using two-loop
time-like splitting kernels to be consistent with the NLO analysis.
The splitting functions was calculated in Refs.~\cite{Stratmann:1996hn} and are
implemented in HOPPET~\cite{Salam:2008qg,Salam:2008sz} which we use in the analysis.
The quality of the agreement between experimental measurements and the
corresponding theoretical predictions for a given set of 
fragmentation parameters is quantified by the log-likelihood function ($\chi^2$), which is given by~\cite{1709.04922}
\begin{equation}
  \chi^2 (\{\alpha,\beta,a_n\}, \{\lambda\}) = \sum_{k = 1}^{N_{\tmop{pt}}} \frac{1}{s_k^2} 
  \left( D_k - T_k - \sum_{\mu = 1}^{N_{\lambda}} \sigma_{k, \mu}
  \lambda_{\mu} \right)^2 + \sum_{\mu = 1}^{N_{\lambda}}
  \lambda_{\mu}^2.
\label{eq:chi2}
\end{equation}
$N_{\tmop{pt}}$ is the number of data points, $s^2_k$ is the total
uncorrelated uncertainties by
adding statistical and uncorrelated systematic uncertainties in quadrature,
$D_k$ is the central value of the experimental measurements, and $T_k$ is the
corresponding theoretical prediction which depends on $\{\alpha,\beta,a_n\}$.
$\sigma_{k, \mu}$ are the correlated errors from source $\mu$ ($N_{\lambda}$ in total).
We assume that the nuisance parameters $\lambda_{\mu}$ follow a standard
normal distribution.
By minimizing $\chi^2 (\{\alpha,\beta,a_n\}, \{ \lambda \})$ with respect to the
nuisance parameters, we get the profiled $\chi^2$ function
\begin{equation}
  \chi^2 (\{\alpha,\beta,a_n\},\{\hat{\lambda}\}) = \sum_{i, j = 1}^{N_{\tmop{pt}}} (T_i - D_i) [\tmop{cov}^{-
  1}]_{ij}  (T_j - D_j),
\end{equation}
where $\tmop{cov}^{- 1}$ is the inverse of the covariance matrix
\begin{equation}
\label{eq:covmat}
  (\mathrm{cov})_{ij} \equiv s_i^2 \delta_{ij} + \sum_{\mu =
  1}^{N_{\lambda}} \sigma_{i, \mu} \sigma_{j, \mu} .
\end{equation}
We neglect correlations of experimental uncertainties between different data points since they are not available. 
However, we include theoretical uncertainties into the covariance matrix of Eq.~(\ref{eq:covmat}) by default, assuming these to be fully correlated among points in each subset of the data shown in Table.~\ref{Tab:data_summary}.
Those are data points within the same bin of the transverse momentum of either the photon, $Z$ boson or jets in the measurement.  
The theoretical uncertainties $\sigma_{j,\mu}$ are estimated by the half width of the scale variations from the prescription mentioned in Sec.~\ref{sec:photon}. 
The best-fit of fragmentation parameters are found via minimization of the $\chi^2$ and further validated through a series of profile scans on each of those parameters.
The scans of the parameter space are carried out with MINUIT~\cite{minuit} program.
We use the text-book criterion of $\Delta\chi^2=1$ on determination of
parameter uncertainties.
It should be noted that tolerance conditions are usually applied for fits involving multiple data sets~\cite{1709.04922} and will lead to conservative estimation of uncertainties.
In addition, we adopt the iterative Hessian approach~\cite{hessian} to generate error sets of
fragmentation functions that can be used for propagation of parameter uncertainties to physical observable.

\subsection{Results and discussions}

The overall agreement between NLO predictions from our nominal fit and the experimental data can be seen from Table.~\ref{tab:dataset}.
The total $\chi^2$ is $267.4$ for a total number of data points of 318.
For the ATLAS dijet measurements which contain the majority of the data points,
the agreement is quite good with $\chi^2/N_{pt}$ well below 1.
Description of the isolated-photon measurements is reasonable with
$\chi^2/N_{pt}\sim 2$.
The agreement to CMS $Z$-boson measurement is good while it is much worse for the ATLAS measurement with $\chi^2/N_{pt}\sim 5$.
The discrepancies to data are mostly due to the low-$p_{T,h}$ kinematic bins ($\sim 4\, {\rm GeV}$) as shown in Appendix~\ref{app:compare}.
For comparison we also include results from alternative fits with either excluding to the theoretical uncertainties or using LO matrix elements and LO evolution of the fragmentation functions.
Impact of the theoretical uncertainties is mostly pronounced for the CMS isolated-photon and $Z$-boson measurements.
The LO fit shows a total $\chi^2$ of more than 3000 indicating the necessity of inclusion of NLO corrections.

\begin{table}[]
  \centering
\begin{tabular}{|c|c|c|c|c|}
\hline
\textbf{Experiments} & \textbf{$N_{pt}$} & \textbf{$\chi^2(/N_{pt})$, NLO} & \textbf{$\chi^2(/N_{pt})$, NLO$_{w/o\,\,th.}$} & \textbf{$\chi^2(/N_{pt})$, LO$_{w/o\,\,th.}$} \\ \hline
CMS $\gamma$ & 5    & 11.3(2.27)      & 28.8(5.76)     & 48.5(9.71)       \\ \hline
ATLAS $\gamma$  & 7      & 17.8(2.55)  & 18.8(2.68)          & 40.5(5.78)  \\ \hline
CMS $Z$ & 11            & 16.2(1.47)  & 24.8(2.25) & 906.9(82.4) \\ \hline
ATLAS $Z$   & 9           & 47.5(5.27)  & 48.1(5.34) & 348.8(38.8)           \\ \hline
ATLAS central jets      & 141      & 98.1(0.69)  & 112.9(0.79)           & 833.7(5.83)   \\ \hline
ATLAS froward jets      & 141        & 76.4(0.53)     & 98.0(0.68)           & 855.6(5.98)   \\ \hline
Total     & 318        & 267.4(0.84)     & 331.2(1.04)           & 3034.0(9.54)   \\ \hline
\end{tabular}
	\caption{
	The $\chi^2$ of individual data sets and their sum from our nominal NLO fit and alternative fits at NLO and LO without theoretical uncertainties.
	Numbers in parenthesis correspond to $\chi^2$ divided by the number of data points.  
	}
  \label{tab:dataset}
\end{table}

The values of all 10 parameters of the fragmentation functions from our nominal best-fit are collected in Table.~\ref{Tab:fit_param}.
In addition we also calculated the first moment of the quark and gluon fragmentation functions $\langle x\rangle$ which corresponds to the total momentum fraction carried by charged hadrons at the initial scale.
The values are 58.6\% and 51.0\% for quark and gluon respectively.
We also show the estimated uncertainties of the fitted parameters in Table.~\ref{Tab:fit_param} as from both the profile scans and the Hessian calculation.
In the latter case, two error sets are generated for each of the 10 orthogonal Hessian directions, and the full uncertainties are obtained by adding uncertainties from individual directions in quadrature~\cite{1709.04922}.
We find good agreements between uncertainties from the two methods in general.
The relative uncertainties of parameters for gluon are $2\sim 5$ times larger than those of the quark, and also are more asymmetric, indicating a larger fraction of quark jets than gluon jet in those measurements.

\begin{table}[]
  \centering
\begin{tabular}{|c|cccccc|}
\hline
 \textbf{quark} & \textbf{$\alpha$} & \textbf{$\beta$} & \textbf{$a_0$} & \textbf{$a_1$}  & \textbf{$a_2$} & \textbf{$\langle x\rangle$} \\ \hline
best-fit & 0.375    & 2.166      & 6.016     & -2.292 &2.083 &0.586    \\ \hline
unc.(scan) & $_{-0.03}^{+0.03}$    & $_{-0.12}^{+0.11}$      & $_{-0.56}^{+0.55}$     & $_{-0.10}^{+0.10}$ &$_{-0.20}^{+0.18}$ &--      \\ \hline
unc.(Hessian) & $_{-0.03}^{+0.03}$    & $_{-0.10}^{+0.09}$      & $_{-0.44}^{+0.45}$     & $_{-0.08}^{+0.08}$ &$_{-0.16}^{+0.16}$ & $_{-0.008}^{+0.007}$     \\ \hline
 \textbf{gluon} & \textbf{$\alpha$} & \textbf{$\beta$} & \textbf{$a_0$} & \textbf{$a_1$}  & \textbf{$a_2$} & \textbf{$\langle x\rangle$} \\ \hline
 best-fit & 0.710    & 10.224      & 44.080     & -3.527 &11.786 &0.510    \\ \hline
unc.(scan) & $_{-0.16}^{+0.09}$    & $_{-0.91}^{+1.09}$      & $_{-13.54}^{+19.54}$     & $_{-0.85}^{+0.95}$ &$_{-3.60}^{+3.54}$ &--     \\ \hline
unc.(Hessian) & $_{-0.10}^{+0.09}$    & $_{-0.93}^{+0.91}$      & $_{-14.1}^{+18.9}$     & $_{-0.83}^{+0.92}$ &$_{-3.52}^{+3.32}$ & $_{-0.012}^{+0.011}$     \\ \hline
\end{tabular}
	\caption{
	The best-fit parameters for quark and gluon from our nominal NLO fit
	and their uncertainties (68\% C.L.) estimated using profile scans or Hessian method.
	The last column is the first moment of the fragmentation functions at the initial scale as calculated using the fitted functional forms.
	}
  \label{Tab:fit_param}
\end{table}

\begin{figure}[htbp]
  \centering
  \includegraphics[width=0.7\textwidth]{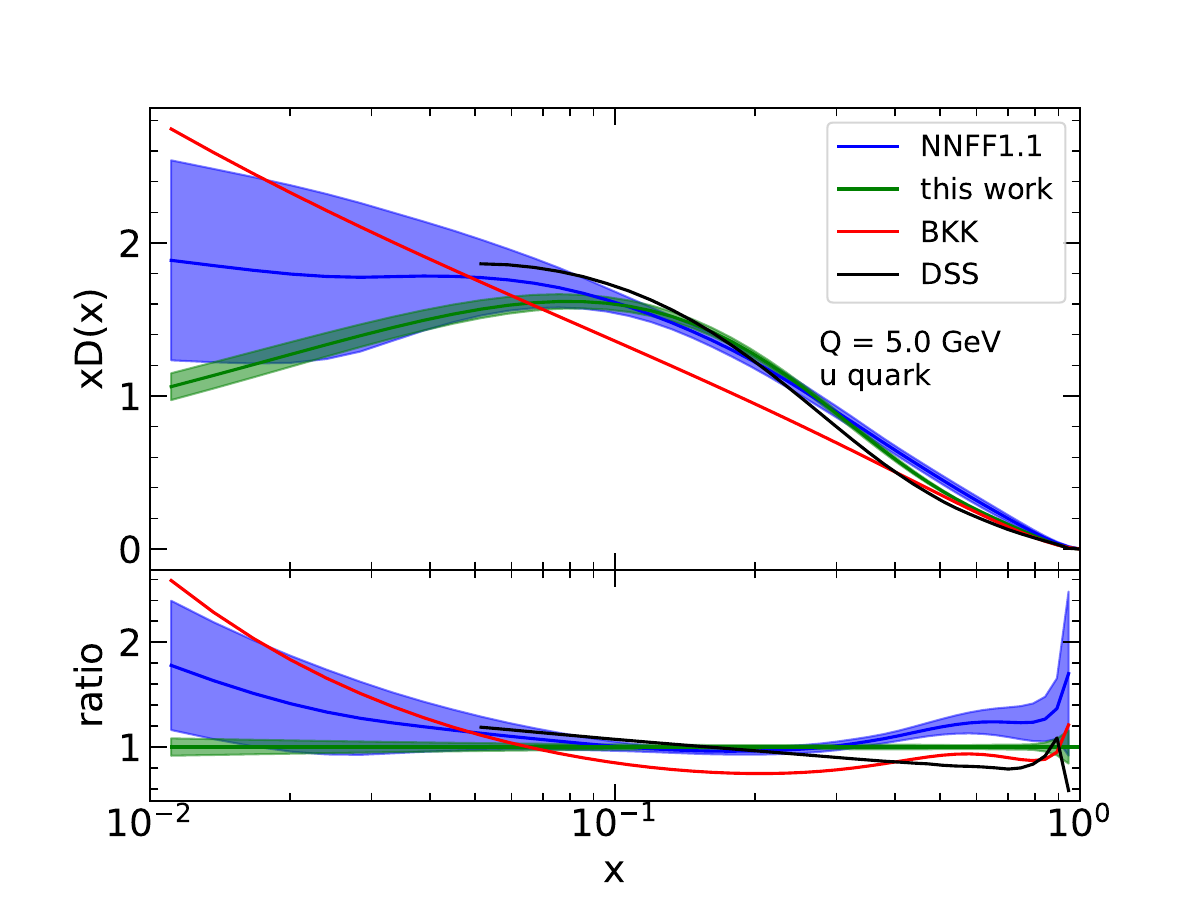}
	\caption{
	The $u$-quark fragmentation function at $Q_0=5$ GeV from our nominal NLO fit as a function of the momentum fraction $x$, and its comparison to the NNFF1.1, BKK and DSS results.
	The colored bands indicate the uncertainties as estimated with the Hessian (MC) method for our (NNFF1.1) fit. 
	}
  \label{Fig:ff-u}
\end{figure}

\begin{figure}[htbp]
  \centering
  \includegraphics[width=0.7\textwidth]{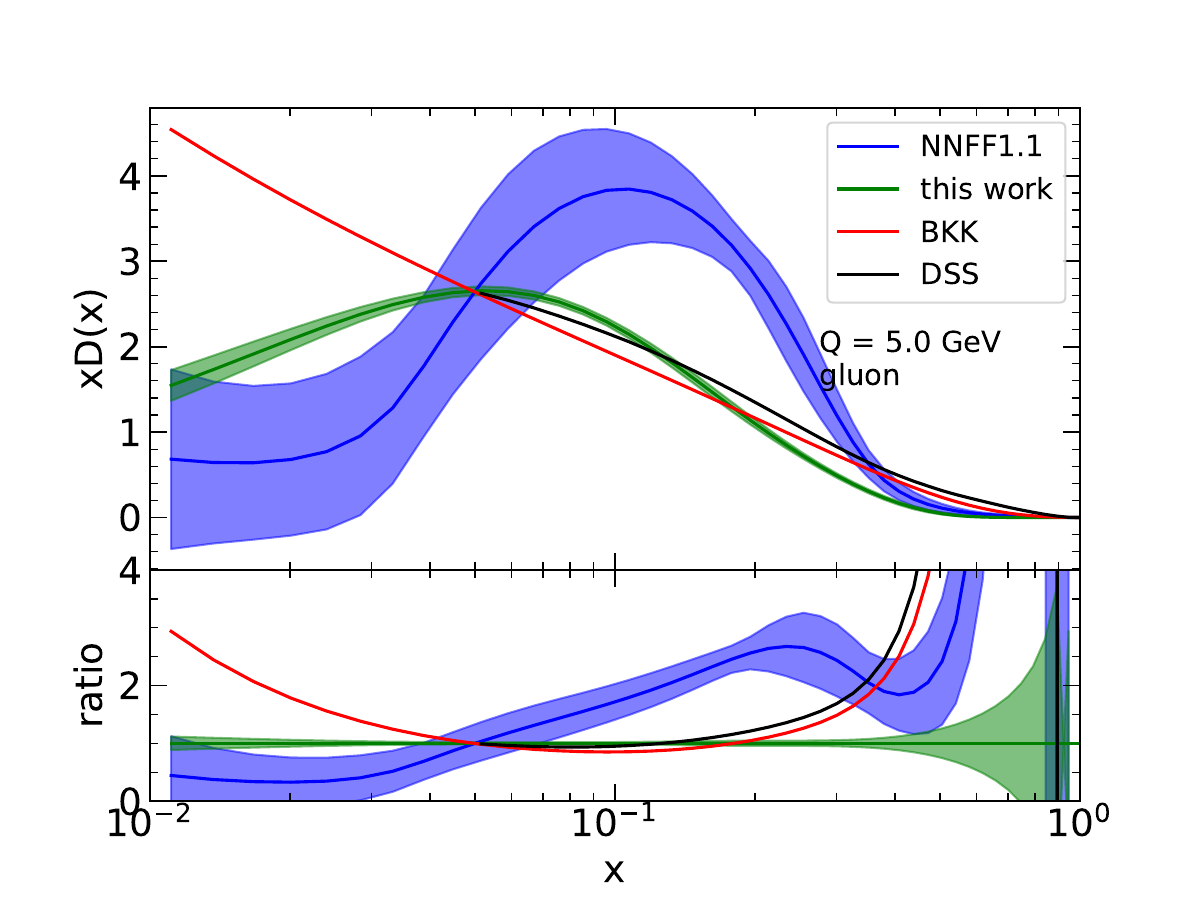}
	\caption{
	The gluon fragmentation function at $Q_0=5$ GeV from our nominal NLO fit as a function of the momentum fraction $x$, and its comparison to the NNFF1.1, BKK and DSS results.
	The colored bands indicate the uncertainties as estimated with the Hessian (MC) method for our (NNFF1.1) fit. 
	}
  \label{Fig:ff-g}
\end{figure}

We compare our fragmentation functions fitted at NLO to those from NNFF1.1, BKK and DSS\cite{dss1,dss2,dss3,dss4,dss5} as a function of the momentum fraction $x$ and at the scale $Q_0=5$ GeV, which is presented in Fig.~\ref{Fig:ff-u} for $u$-quark and in Fig.~\ref{Fig:ff-g} for gluon respectively.
We should emphasize that in this comparison of our fit we use a restricted parametrization form, namely setting equal of all quark fragmentation functions, which are allowed to be different in fits of NNFF1.1, BKK and DSS, and the error criterion chosen by us is $\Delta \chi^2=1$. 
%{\alert Therefore,  the small uncertainties in our work does not necessarily indicate better constrained analysis but stems from a specific assumption.}
It should be noted that the small uncertainties in our work, which will be shown below, are partly attributed to the specific assumption we have made in our parametrization.

The upper panel in both figures illustrates the value of the fragmentation function multiplied by the momentum fraction $x$ as a function of $x$ and in the lower one all results are normalized to the central value of our findings. Finally, we need to mention that the colored bands represent the estimated uncertainties from the corresponding fits when they are available.
%
% While the error band in our work appears to be small, it should be noted that we have made the assumption of no separation of quark flavors, and the error tolerance chosen by us is $\Delta \chi^2=1$.
%

%

For the $u$ quark, 
we observe a good agreement between NNFF1.1 results and our work in the region $0.1 < x < 0.3$ in Fig.~\ref{Fig:ff-u}. 
However, significant deviations, especially in the small $x$ region, are observed. These deviations can reach up to nearly 80\%. 
On the other hand, our results remain within the error band of NNFF1.1 results in lower regions. 
It is important to note that the error band of NNFF1.1 data is large in the small $x$ region (where $x < 0.1$), indicating a lack of well-fitted data in that range.
Comparatively, the BKK results exhibit significant deviations from our findings throughout the entire region. 
Particularly in the small $x$ region, the deviation can be as large as 160\%. Unfortunately, the error data for BKK is not available.
DSS data only fits from 0.05 to 1, and errors are not provided. In the region of $0.05<x<0.3$, a close resemblance with our work is also observed with the maximum deviation about 20\%.
Furthermore, both the BKK and NNFF1.1 data show a decreasing trend as $x$ increases, DSS data, in its available region, a decreasing trend is also observed, while our results demonstrate an increase as $x$ approaches approximately 0.1, followed by a decrease.

In the case of the gluon, Fig.~\ref{Fig:ff-g} reveals significant discrepancies among the four results. Firstly, it is evident that these results do not agree except for the BKK and DSS fits, indicating a tension between different fits of the gluon fragmentation function. Secondly, the error band associated with NNFF1.1 data is notably large, suggesting a higher level of uncertainties in that dataset.
Further examination reveals that in certain regions (specifically for $x<0.5$), the ratio with respect to our work is less than 4. However, beyond this range, the ratio experiences a sudden and pronounced increase.
The upper panel also highlights the contrasting trends exhibited by the BKK and DSS results. Specifically, the BKK results consistently decrease as $x$ increases. But DSS results are not available at $0.01<x<0.05$, therefore, its property is not known in these regions.
In contrast, our fit and the NNFF1.1 data initially display an increase, followed by a subsequent decrease. Our results reach their maximum value at approximately $x=0.05$, while the NNFF1.1 data reach their peak at around $x=0.1$.
The notable disparities among different datasets emphasize the need for additional constraints and further data to improve the accuracy of the gluon fit.

\section{Conclusions}
\label{sec:conclusions}

In this work, we propose a new prescription for combining general-purpose Monte-Carlo generators with fragmentation functions (FFs) at NLO in QCD.
This new framework, dubbed FMNLO, is based on a hybrid scheme of NLO calculations utilizing a phase-space slicing of collinear regions in combination with the usual local subtraction methods, 
and organizes various ingredients for fragmentation predictions in a way suitable for Monte Carlo calculations.
As a proof of concept, we realize FMNLO with \MG. 
The corresponding code is publicly available and is introduced in Appendix~ \ref{app:install}.
Our scheme and its implementation are validated 
for several scenarios in both lepton collisions and hadron collisions. 
The combination of general-purpose MC generators and FFs allows for the 
study of single-hadron production for various hard process at NLO in QCD 
with general selection cuts or jet reconstruction. 
As examples, we compare the predictions of FMNLO with experimental measurements of jet production with a tagged isolated photon, 
jet production with a tagged $Z$ boson, and inclusive dijet production.
Also, we boost FMNLO with interpolation techniques, 
such that for a given measurement, the time-consuming calculation
of NLO partonic cross section can be 
%done for only once, 
reused when the fragmentation functions are changed.
The combination of these two features endows it with the unique ability of 
making theoretical predictions for a wide range of measurements within a reasonable time consumption, which is essential for a global fit of FFs.
We demonstrate this ability by performing a NLO fit of parton FFs to unidentified 
charged hadrons, using hadron production measurements at the LHC. 
Our nominal fit shows very good agreements with the LHC data.
We find that the high-precision fragmentation measurements from ATLAS inclusive dijet production especially show a strong constraining power on the FFs.
Our unidentified charged-hadron FFs are then compared with those from BKK, DSS and NNFF1.1. 
Notable disparity in gluon FF is found, indicating the necessities of
additional constraints and data in gluon fit.
We emphasize that our framework also works for FFs with specific flavors. 

Besides its ability in extraction of FFs, 
the proposed scheme and its implementation open the opportunity of studying BSM effects with single-hadron production. 
 FMNLO is also desirable for calculations of NLO hard functions needed for various predictions of QCD resummation~\cite{Gao:2019ojf,Chen:2019bpb,Kang:2020yqw,Li:2023dhb}. 
Furthermore, it can be generalized to calculate distributions of observable related to transverse dependent fragmentation functions which have been widely used in studies of jet substructures~\cite{Kang:2019ahe,Luo:2020epw,Chien:2022wiq}. 
We leave those for future investigations and updates of the program.

\section*{Acknowledgments}
We would like to thank HX Zhu for motivating this work as well as contributions at early stage.
JG would like to thank HX Xing for discussions, B. Nachman and D. Kar for communications on the ATLAS dijet data.
This work was sponsored by the National Natural Science Foundation
of China under the Grant No.12275173 and No.11835005.
The work of X. Shen is supported in part by the Helmholtz-OCPC Postdoctoral Exchange Program under Grant No. ZD2022004.

\appendix

\section{Installation and running}
\label{app:install}
The current version of the program \texttt{FMNLOv1.0} and the related publications can be found on the website\footnote{\url{http://fmnlo.sjtu.edu.cn/~fmnlo/}}.
Prerequisites on running the program include \texttt{MG5\_aMC@NLO}~\cite{Alwall:2014hca,Frederix:2018nkq} and possibly \texttt{LHAPDF6} library~\cite{lhapdf6} for invoking fragmentation functions therein.  
The recommended version of \texttt{MG5\_aMC@NLO} is \texttt{3.4.0} which has been tested thoroughly.
The package \texttt{FastJet}~\cite{Cacciari:2005hq,Cacciari:2011ma} is also required which can be installed within \texttt{MG5\_aMC@NLO}.
The paths to \texttt{MG5\_aMC@NLO} and \texttt{LHAPDF6} can be set separately at the top of \texttt{Makefile} under the main directory \texttt{FMNLOv1.0} and of \texttt{mgen.sh} under directory \texttt{mgen}.
The \texttt{HOPPET v1.2.1-devel}~\cite{Salam:2008qg,Salam:2008sz} program has been modified and integrated into the source file. 
Users should cite original works of those external programs properly together with this publication.
One can simply run \texttt{make} under the main directory to compile all ingredients.
Note that in order to retain full quark-flavor information in \texttt{MG5\_aMC@NLO}, one has to modify \texttt{madgraph/core/helas\_objects.py} in the native \texttt{MG} directory to disable group of identical processes. 
That can be achieved by simply replace \texttt{True} with \texttt{False} at the \texttt{3486-th} line of the file.
We mentioned earlier that the FKS subtractions have not been implemented for NLO calculations of DIS processes in \texttt{MG5\_aMC@NLO}.
However, we are working toward an alternative solution that will come with the next release of \texttt{FMNLO}.
To calculate the parton fragmentation in a typical hard scattering process, two subsequent steps are followed.
First, inside the directory \texttt{mgen} one invokes \texttt{MG5\_aMC@NLO} with a customized analysis routine (a \texttt{module}) to generate the interpolation tables storing matrices of coefficient functions in Eq.~\ref{Eq:coe}. %equation 2.11
We have released modules for all processes included in this study.
New modules can be added easily following existing examples.
In \texttt{mgen}, the subdirectory \texttt{common} contains the common ingredients needed for all modules.
Each module has a separate directory including \texttt{init.sh} for \texttt{MG5\_aMC@NLO} command, \texttt{pre\_cuts.f} for the selection of relevant phase space, and \texttt{analysis\_HwU\_pp.f} containing the main analysis routine.
Various input parameters are specified in the file \texttt{proc.run}.
Each line contains a record for one input variable:
a character tag with the name of the variable, followed by the
variable's value.
We take \texttt{proc.run} used for the calculation of CMS isolated-photon-tagged jet measurement as an example.
\renewcommand{\baselinestretch}{0.78}
\begin{verbatim}
    # main input for generation of NLO fragmentation grid file by MG5
    process A180104895
    # subgrids with name tags
    grid pp
    obs 4
    cut 0.02
    pta1 60.0
    pta2 10000.0
    ptj1 30.0
    ptj2 10000.0
    # in MG5 format
    set lpp1 1
    set lpp2 1
    set ebeam1 2510.0
    set ebeam2 2510.0
    set lhaid 13100
    set iseed 11
    set muR_over_ref 1.0
    set muF_over_ref 1.0
    end
\end{verbatim}
\begin{itemize}
\item \texttt{process} specifies the name of the directory that contains the module to be loaded.
\item \texttt{grid} is a string indicating the name of the running job.
\item \texttt{obs} specifies different distributions to be calculated:
\texttt{1} for distribution in $\zeta$, \texttt{2} for distribution in $p_{T,h}$, \texttt{3} and \texttt{4} for distributions in $\xi_{T}^{j}$ or $\xi_{T}^{\gamma(Z)}$.
\item \texttt{cut} gives the slicing parameter $\lambda$ and a value of $0.02$ is recommended.
\item \texttt{pta1} and \texttt{pta2} specify the lower and upper limit of the kinematic range of the transverse momentum of the photon.
\item \texttt{ptj1} and \texttt{ptj2} specify the lower and upper limit of the kinematic range of the transverse momentum of the jet.
\item Other possible inputs in this block include \texttt{hrap} and \texttt{pth} specifying the upper limit on the absolute pseudorapidity and the lower limit on the transverse momentum of the hadrons. \texttt{isca} determines the choice on central value of the fragmentation scale,
\texttt{1} for our nominal choice of $\max\{p_{T,j}\}$ ($Q$) for $pp$ ($e^+e^-$) collisions, and \texttt{2} for using $p_{T,h}$ ($E_h$).
Default values of all above variables are assigned via the script file \texttt{mgen.sh} for individual modules if not specified in \texttt{proc.run}.  
\item The remaining inputs follow the same syntax as the normal \texttt{MG5\_aMC@NLO} command, for instance, \texttt{lpp1} and \texttt{ebeam1} are type and energy of collision particle 1, \texttt{lhaid} specifies parton distribution of proton used for the calculation, etc. 
\end{itemize}
The generation of fragmentation grid can be launched by the command \texttt{./mgen.sh proc.run}.
Note that for the same process, generation of multiple grids can be grouped into a single input file by simply repeating the two blocks
after the \texttt{process} line.
Once the generation of grid is finished, it will be stored in an upper-level directory \texttt{grid}, for instance with a name \texttt{A180104895\_pp.fmg} for above example. 
After generation of the grid, the calculation of physical distributions
can be done within seconds by running \texttt{./fmnlo} in the directory \texttt{data}.
Input parameters at this stage are specified in the file \texttt{input.card}.
\renewcommand{\baselinestretch}{0.78}
\begin{verbatim}
 1  #  loop for D fun (1/2 -> LO/NLO) | evo for D fun (0/1 -> internal/hoppet)
 2  #  followed by >=1/0 -> internal/LHAPDF | FFID | FFmember 
 3  2       0
 4  0       NNFF11_HadronSum_nlo            0
 5  #  normalization | grid file | binnig file 
 6  #  0/1/2 -> absolute dis./normalized to corresponding order/leading order
 7  #  can include multiple entries in several lines
 8  1 "../grid/A180104895_pp.fmg" "../grid/1801-04895.Bin"
\end{verbatim}
\begin{itemize}
\item The \texttt{3rd} line specifies the order of DGLAP evolution of the fragmentation functions, \texttt{1/2} for LO and NLO respectively, followed by \texttt{0/1} for using the native evolution provided by the input fragmentation functions or evolving with \texttt{HOPPET} package from the initial scale $Q_0$.
\item The \texttt{4th} line indicates the choice of fragmentation functions. 
A value of \texttt{0} indicates the usage of fragmentation functions from LHAPDF6 library and other integer values correspond to fragmentation functions implemented in \texttt{FMNLO1.0}, e.g., \texttt{1} for the NLO nominal fragmentation functions presented in this study (note one should use the NLO evolution with \texttt{HOPPET} concurrently), \texttt{2} for the BKK functions of unidentified charged hadrons, and \texttt{3(4)} for the KKP~\cite{Kniehl:2000fe}(DSS) functions.
The following two inputs specify the name and set number of the fragmentation function in the case of using LHAPDF6.
Note that the value of the QCD coupling used in the evolution of fragmentation functions is set consistently. It can either be imported from LHAPDF6 or set by HOPPET with $\alpha_S(M_Z)$ fixed at 0.118.
Other possibilities on the choice of fragmentation function can be implemented by modifying the source file \texttt{internal.f}. 
\item The \texttt{8th} line specifies choice of normalization: \texttt{0} for absolute distributions, \texttt{1} and \texttt{2} for normalized distributions to the total cross sections of corresponding order or LO respectively. 
The followed are the name of the pre-generated grid file for the calculation and the name of the file containing binning of the distribution.
Multiple entries similar to the \texttt{8th} line can be added to calculate several distributions at once.
The binning is set via two-line inputs.
For example, in \texttt{1801-04895.Bin}, 
\renewcommand{\baselinestretch}{0.78}
\begin{verbatim}
  8
  0.5  1.0  1.5  2.0  2.5  3.0  3.5  4.0  4.5
\end{verbatim}
the first line specifies the total number of kinematic bins and the second line contains all nodes of the bins in sequence.  
\end{itemize}
Once \texttt{./fmnlo} is executed, the format of the printout can be understood easily 
\renewcommand{\baselinestretch}{0.78}
\begin{verbatim}
   ID(1/x dx/dkv)     zd     zu    LO*{1,0.5,2}   NLO*{1,0.5,2}  NLO/LO
   1   5.00000E-01   1.00000E+00   2.45041E-01 ..  2.61256E-01 ..  1.066 ..
   2   1.00000E+00   1.50000E+00   6.69355E-01 ..  7.61367E-01 ..  1.137 ..
   3   1.50000E+00   2.00000E+00   1.33705E+00 ..  1.61524E+00 ..  1.208 ..
   4   2.00000E+00   2.50000E+00   2.12904E+00 ..  2.51633E+00 ..  1.182 ..
   5   2.50000E+00   3.00000E+00   2.93954E+00 ..  3.23861E+00 ..  1.102 ..
   6   3.00000E+00   3.50000E+00   3.65409E+00 ..  3.60505E+00 ..  0.987 ..
   7   3.50000E+00   4.00000E+00   4.21952E+00 ..  3.12574E+00 ..  0.741 ..
   8   4.00000E+00   4.50000E+00   2.98053E+00 ..  1.44477E+00 ..  0.485 ..
\end{verbatim}
which contains the distribution at LO and NLO for three choices of the fragmentation scale $\mu_D=\{1,1/2,2\}\mu_{D,0}$ and the ratio of NLO to LO predictions for all kinematic bins specified.

\section{Comparison of the theory to data}
\label{app:compare}

\begin{figure}[htbp]
  \centering
  \includegraphics[width=0.32\textwidth]{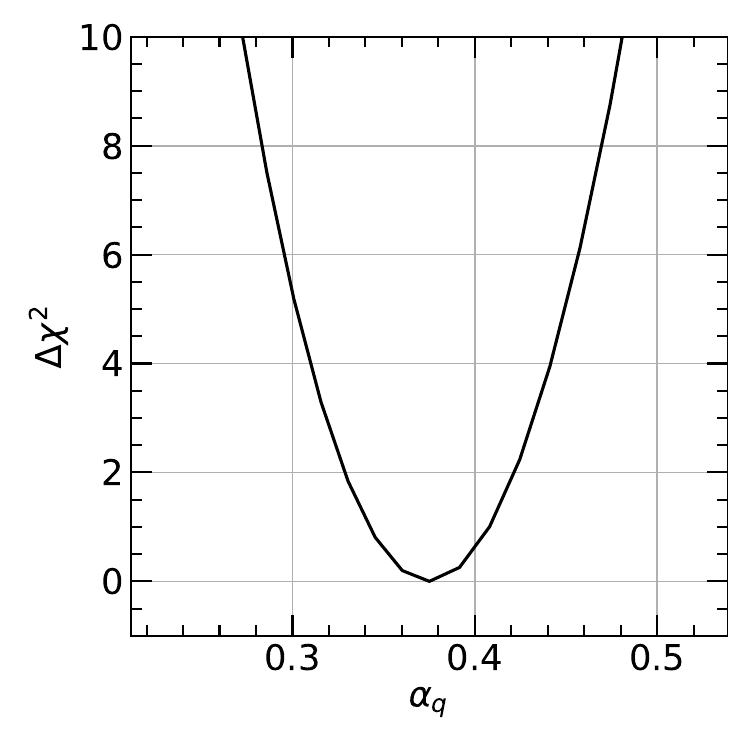}
  \includegraphics[width=0.32\textwidth]{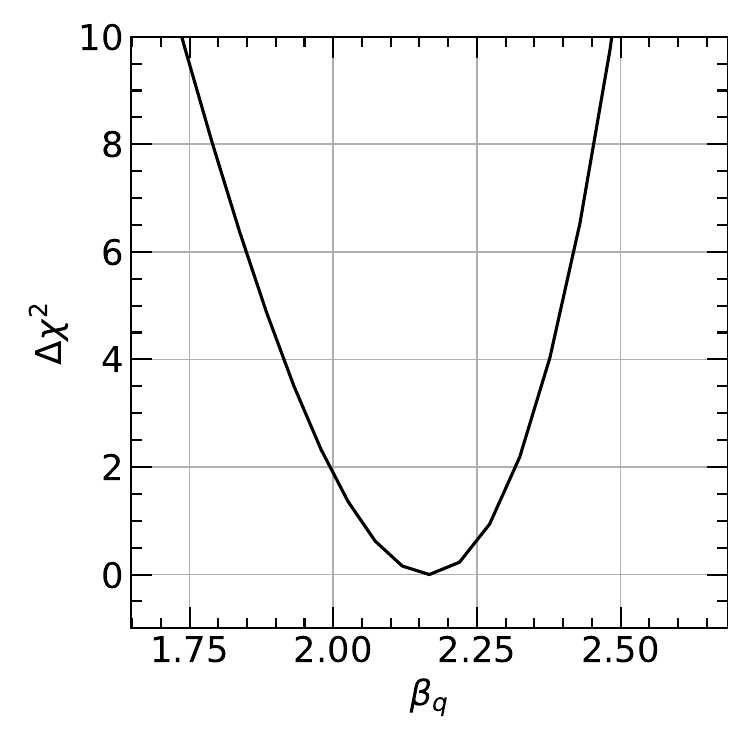}
  \includegraphics[width=0.32\textwidth]{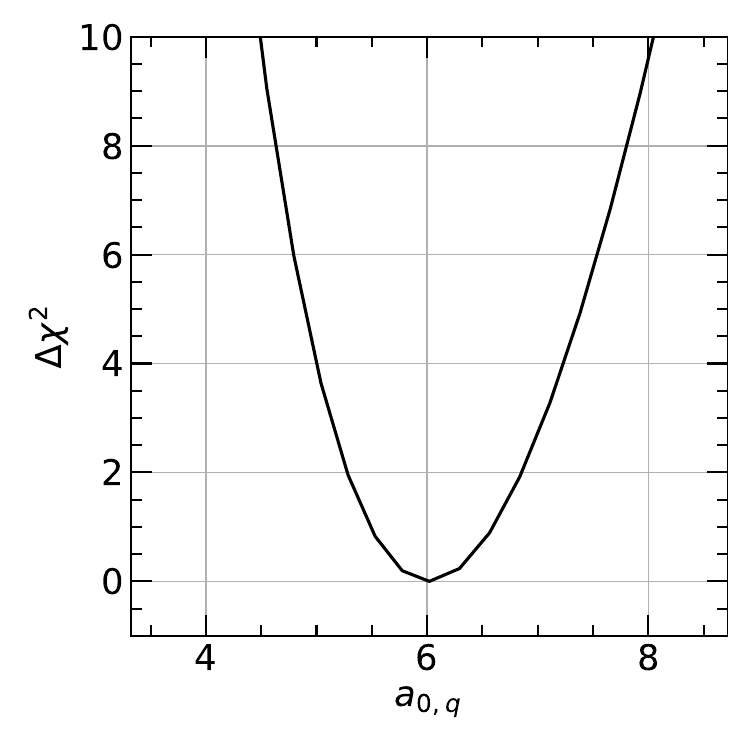}
  \includegraphics[width=0.32\textwidth]{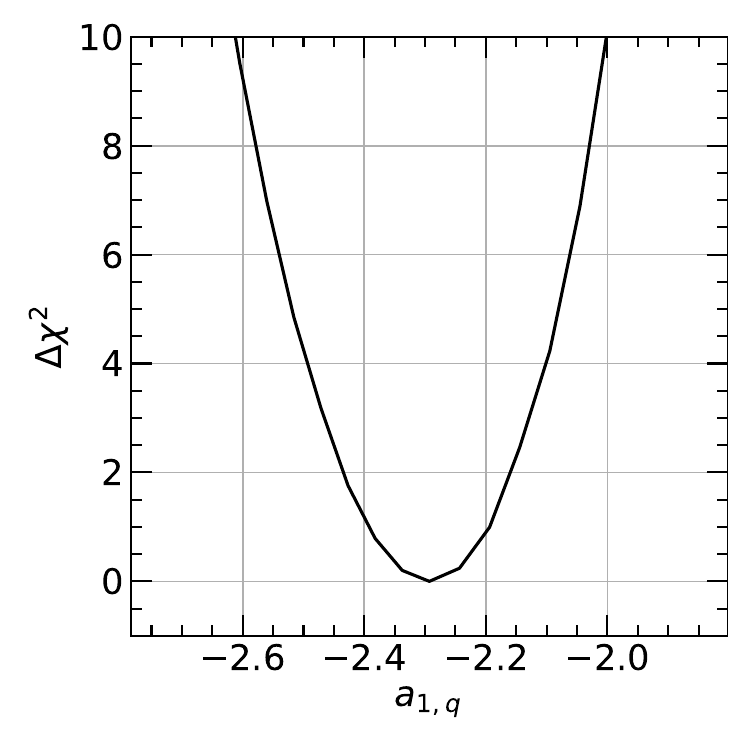}
  \includegraphics[width=0.32\textwidth]{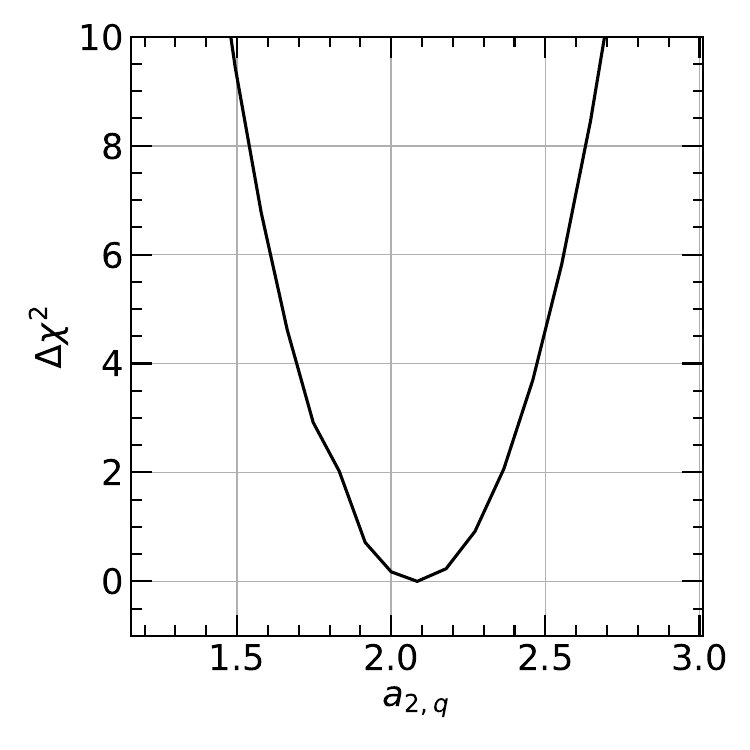}
    \captionsetup{font={stretch=1}}
	\caption{
	Profile of total $\chi^2$ change from scans on individual parameters of the quark fragmentation functions under nominal fit with other parameters freely varying.
	}
  \label{Fig:qpara}
\end{figure}

\begin{figure}[htbp]
  \centering
  \includegraphics[width=0.32\textwidth]{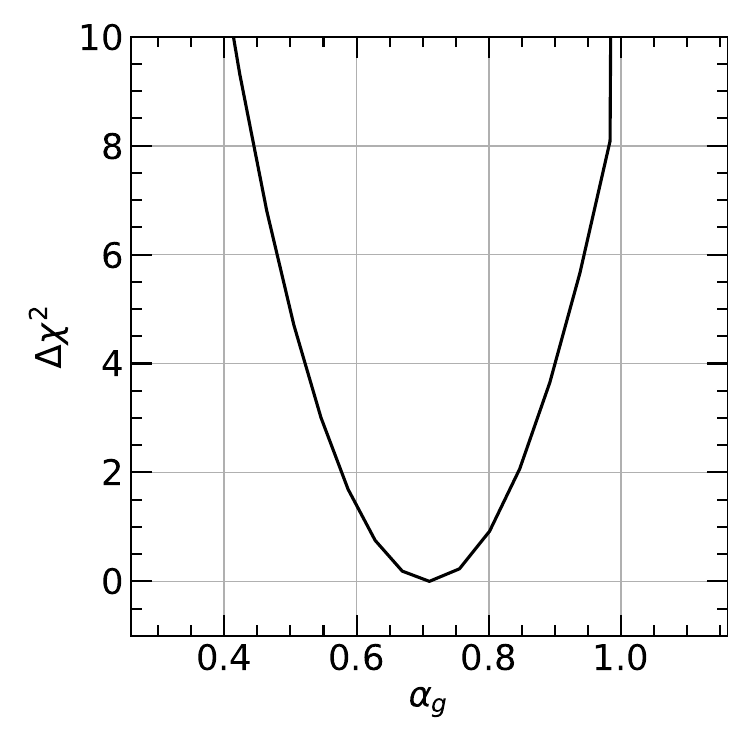}
  \includegraphics[width=0.32\textwidth]{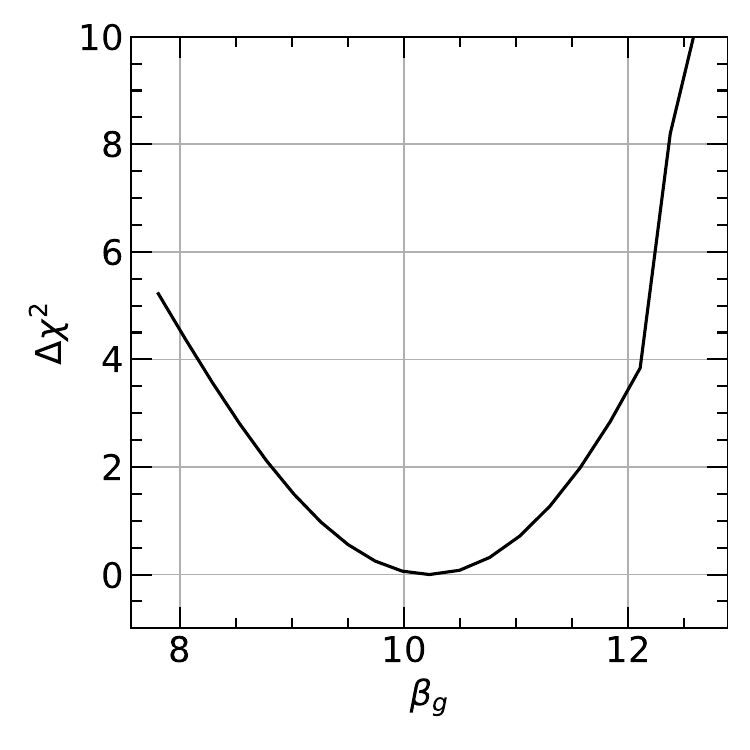}
  \includegraphics[width=0.32\textwidth]{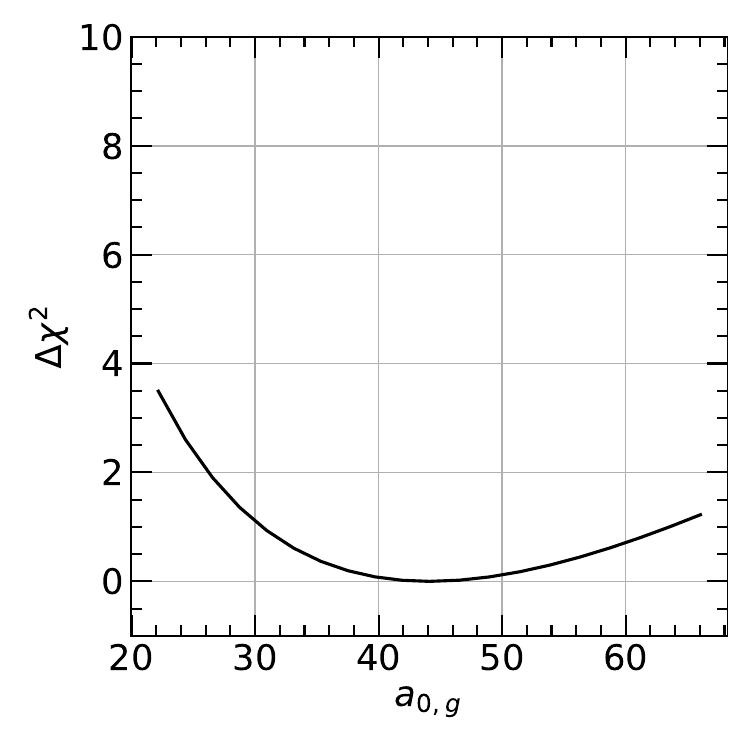}
  \includegraphics[width=0.32\textwidth]{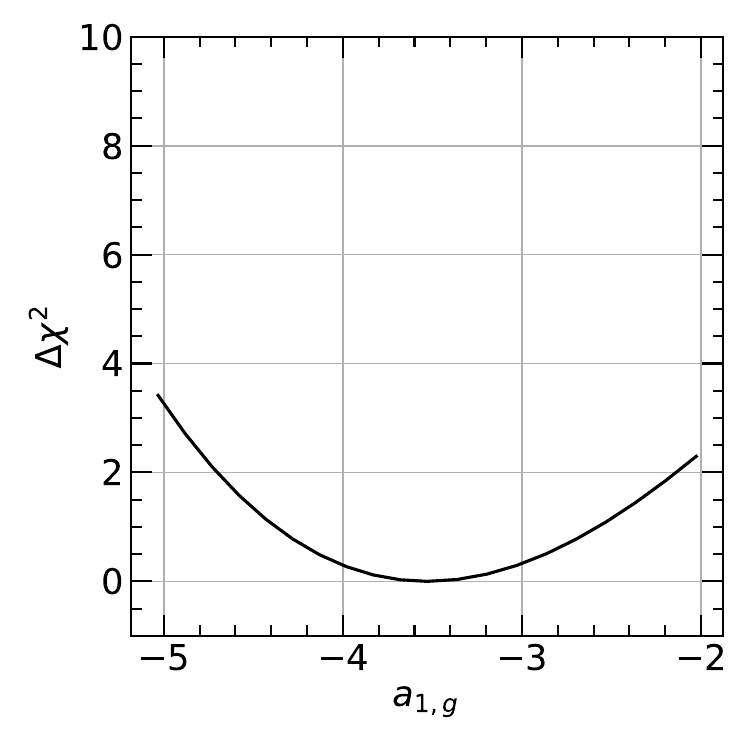}
  \includegraphics[width=0.32\textwidth]{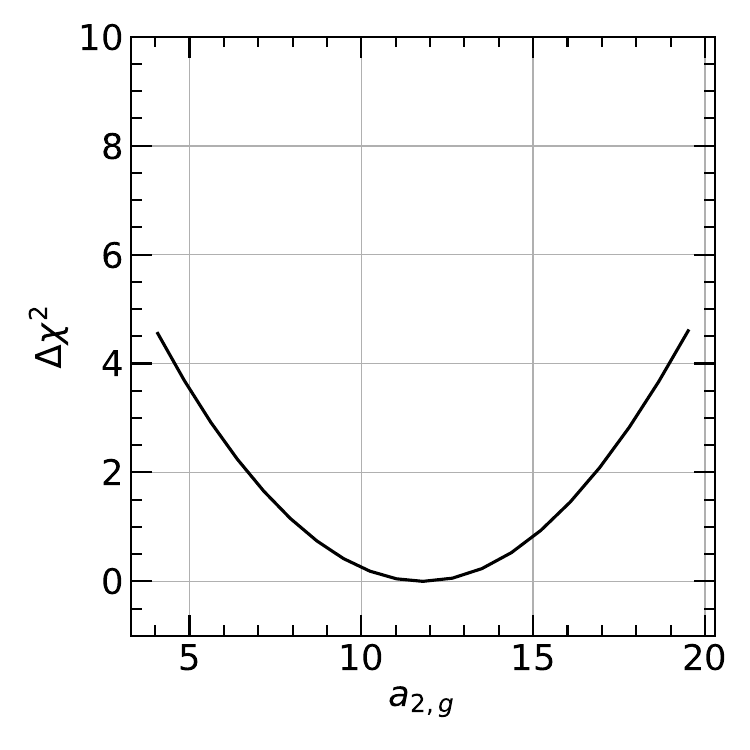}
    \captionsetup{font={stretch=1}}
	\caption{	
	Profile of total $\chi^2$ change from scans on individual parameters of the gluon fragmentation functions under nominal fit with other parameters freely varying.
	}
  \label{Fig:gpara}
\end{figure}

In this appendix we include more details on our nominal NLO fit to the LHC data.
We first show the total $\chi^2$ profile from scans on individual parameters of the fragmentation functions in Fig.~\ref{Fig:qpara} and~\ref{Fig:gpara} for quark and gluon respectively.
In each scan all other parameters are set free and are fitted to minimize the constrained $\chi^2$. 
The fragmentation function of quarks is better constrained as mentioned above.
The $\chi^2$ profile shows a parabolic shape around the minimum.
The fragmentation function of gluon is less constrained especially for parameters $a_{0/1/2}$.
For instance, the increase of $\chi^2$ can barely reach 2 units even we scan over a wide range of $a_{0}$ and $a_{1}$.
The large variations of the parameters also lead to a non-parabolic shape of the $\chi^2$ profile in the scan region. 
\begin{figure}[htbp]
  \centering
  \includegraphics[width=0.43\textwidth]{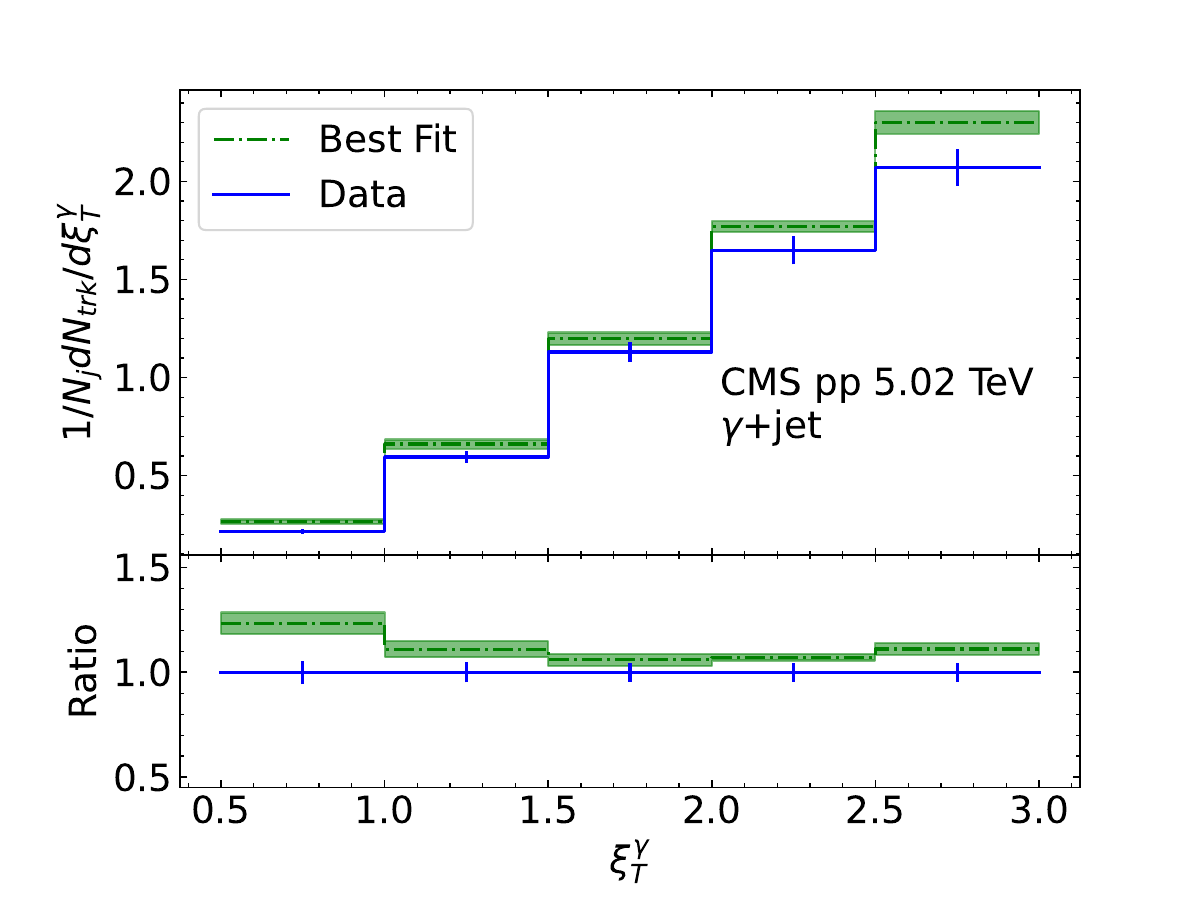}
  \includegraphics[width=0.43\textwidth]{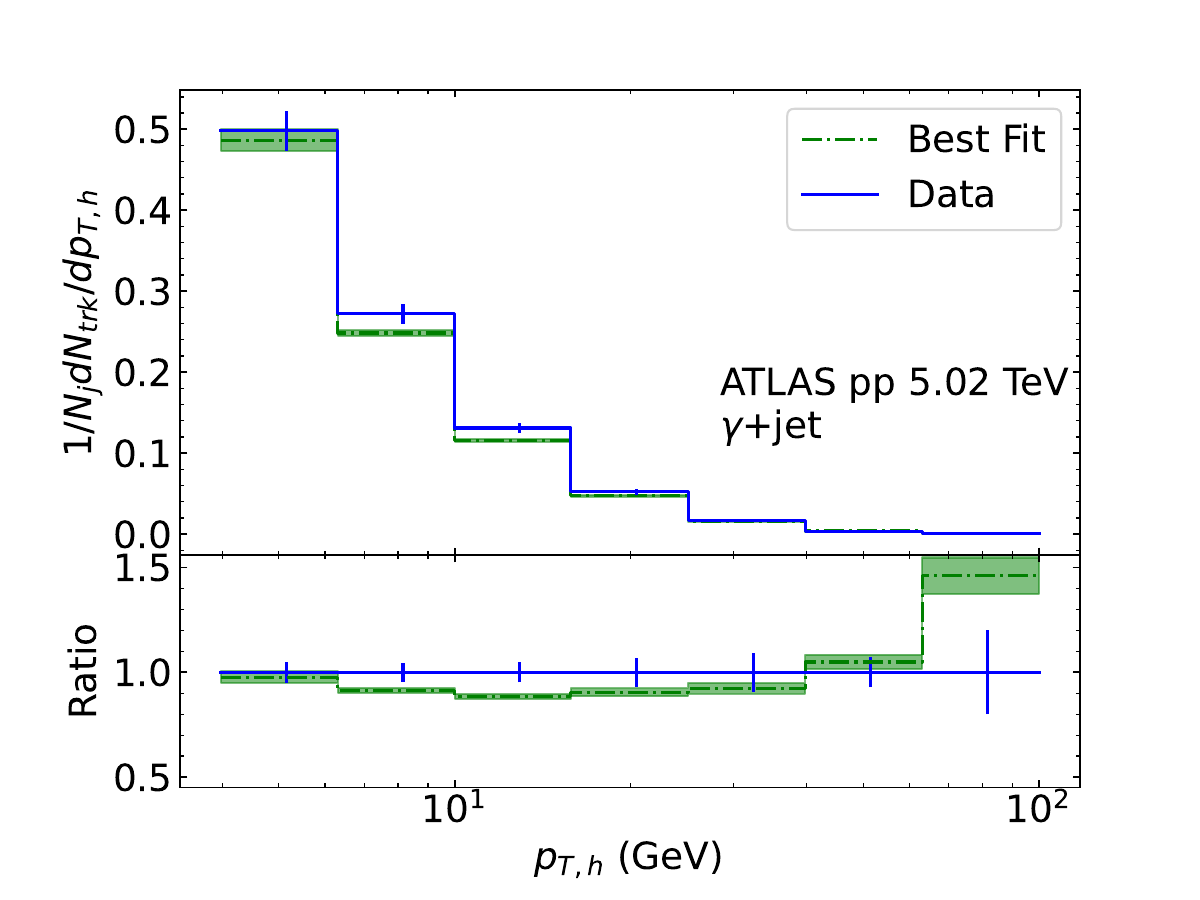}
    \captionsetup{font={stretch=1}}
	\caption{
	Our NLO predictions, obtained using the best-fit fragmentation functions, are shown and compared to the CMS and ATLAS data on the isolated photon production. 
	The blue solid line and green dash-dotted line represent experimental data and our best-fit results respectively.
	The enveloped scale uncertainty is shown by the colored bands. 
	The total experimental uncertainty is shown by the  error bar.
	The results are normalized to central value of the experimental data and shown in the lower panel with the same colors and line styles.
	}
  \label{Fig:thdagamma}
\end{figure}
\begin{figure}[htbp]
  \centering
  \includegraphics[width=0.43\textwidth]{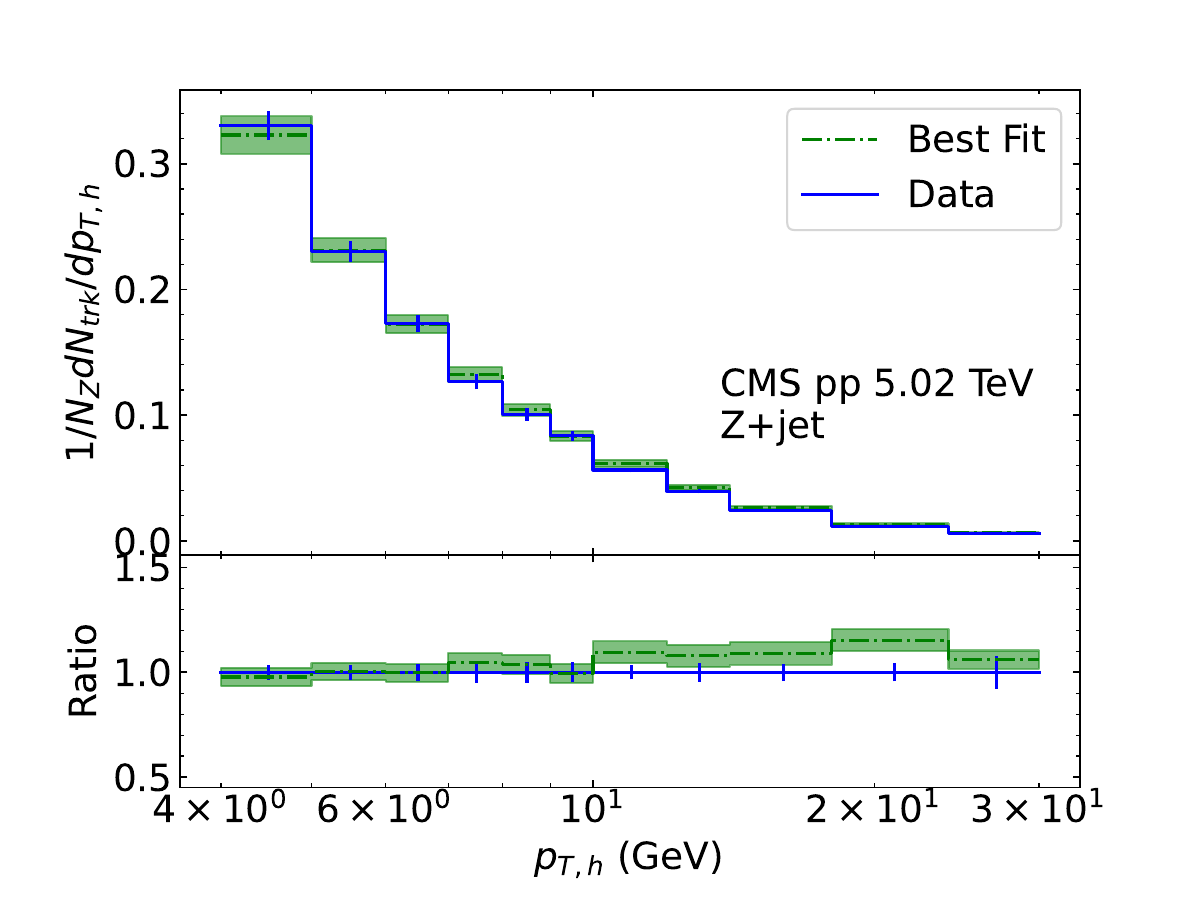}
  \includegraphics[width=0.43\textwidth]{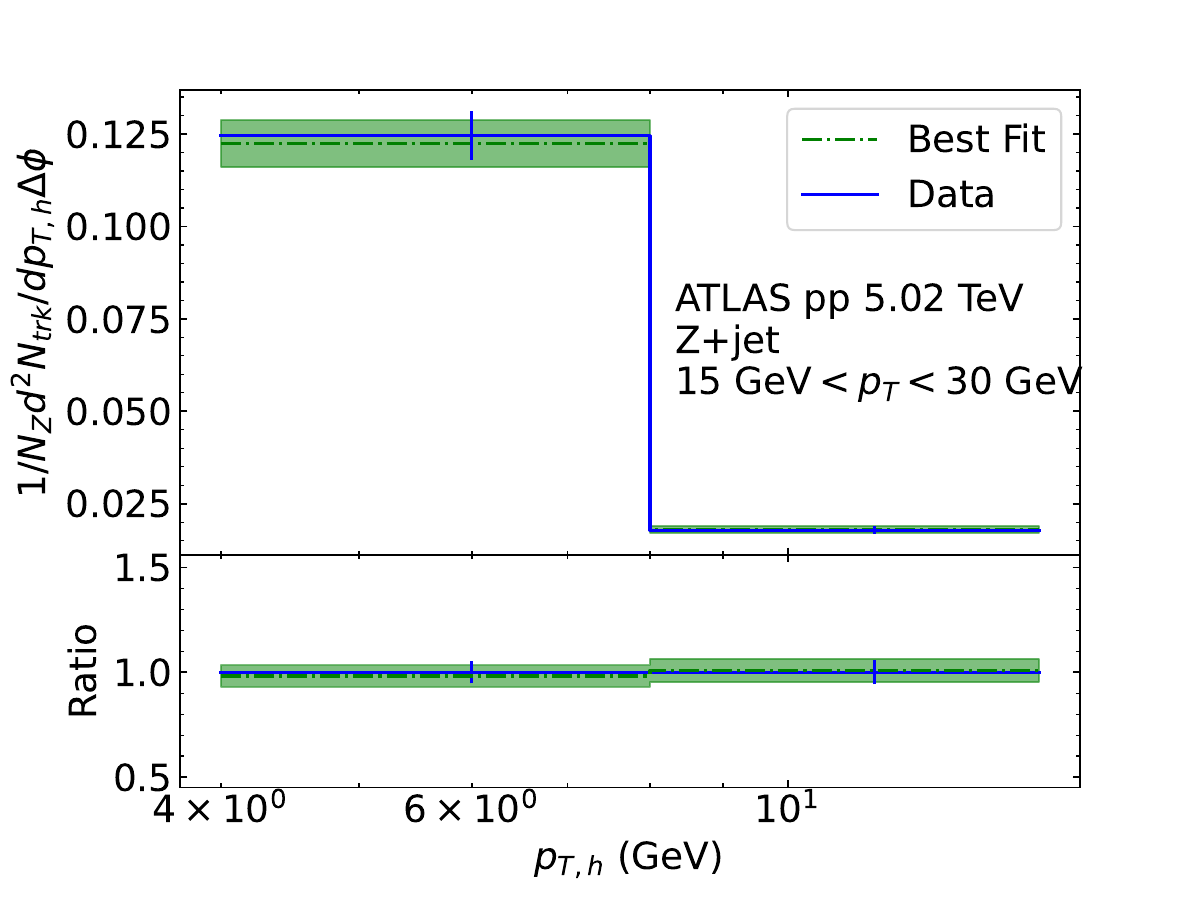}
  \includegraphics[width=0.43\textwidth]{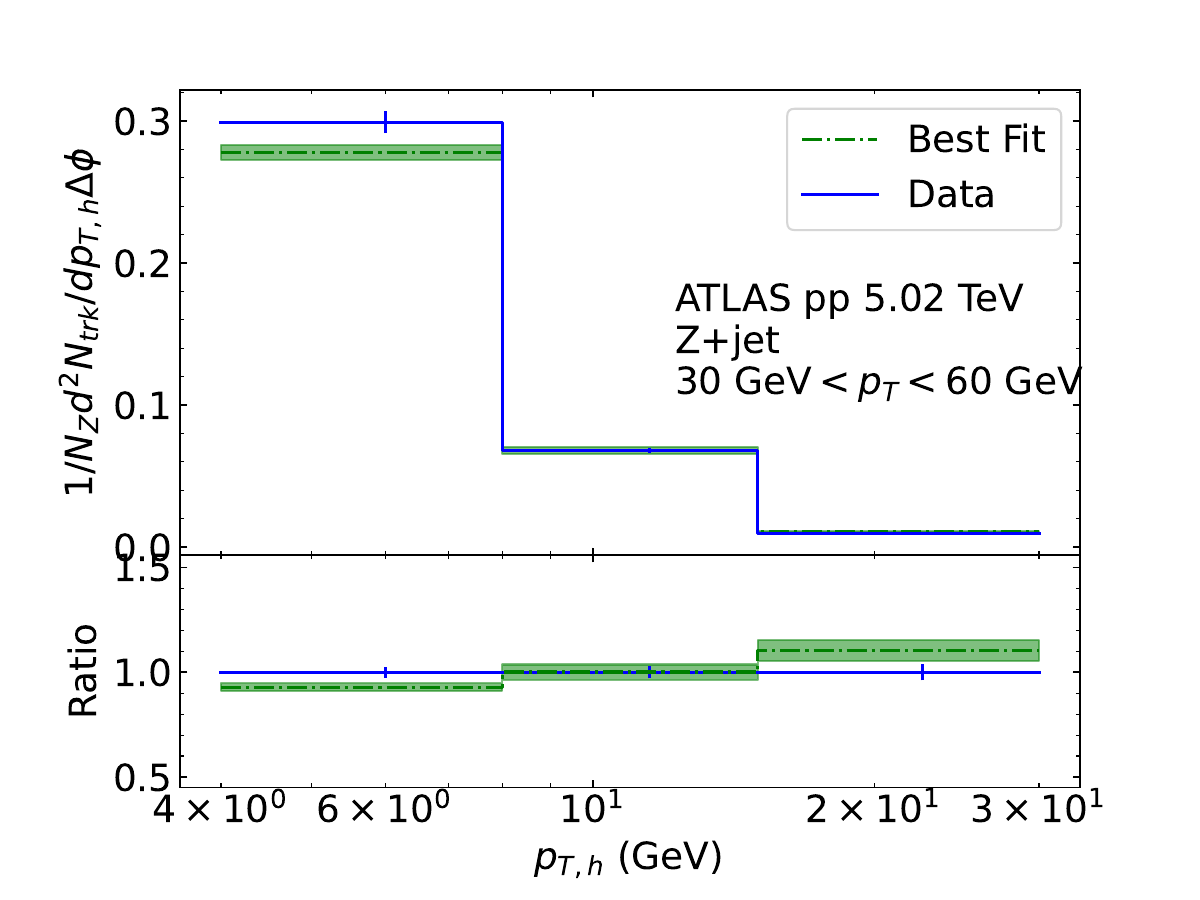}
  \includegraphics[width=0.43\textwidth]{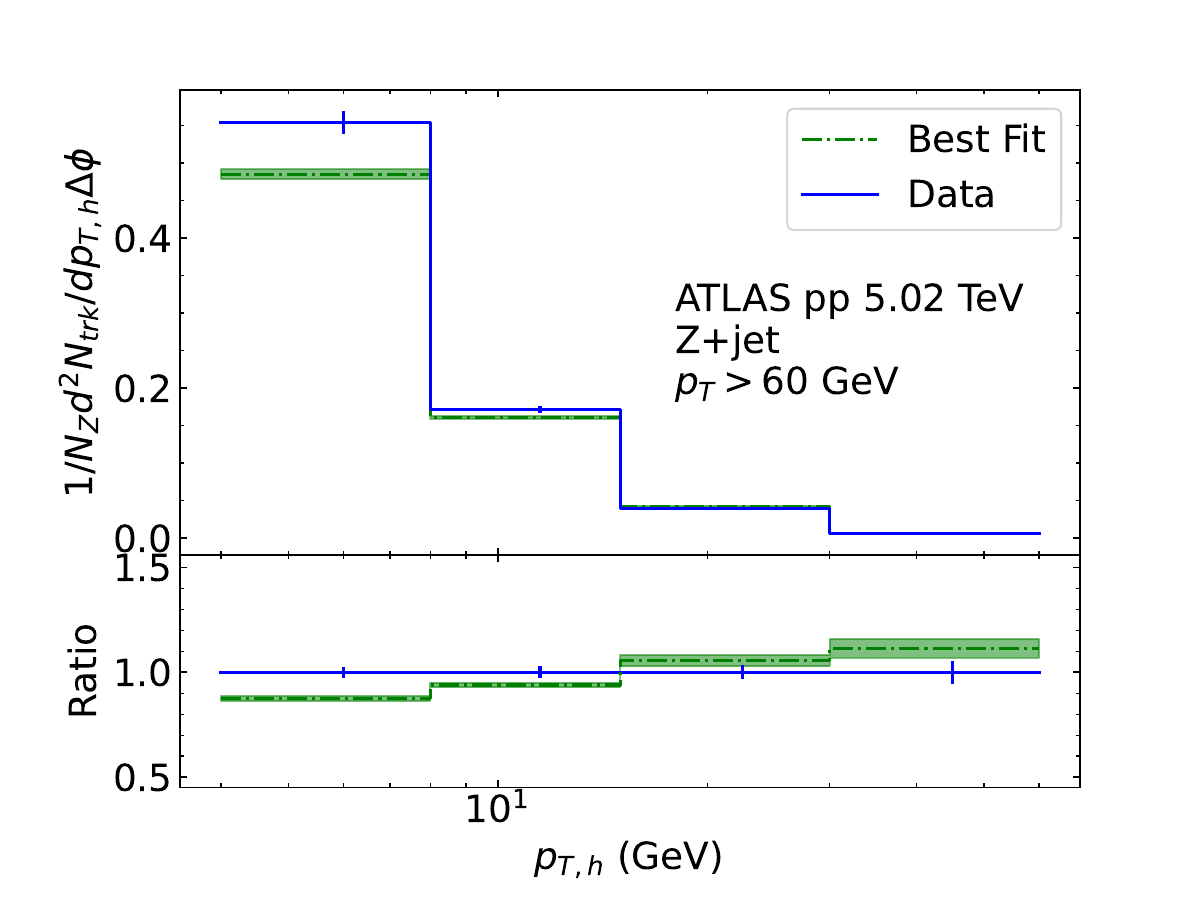}
    \captionsetup{font={stretch=1}}
	\caption{
	Similar to Fig.~\ref{Fig:thdagamma} but for the $Z$ boson production and in different bins of transverse momentum of the $Z$ boson.
	}
  \label{Fig:thdazboson}
\end{figure}
\begin{figure}[htbp]
  \centering
  \includegraphics[width=0.43\textwidth]{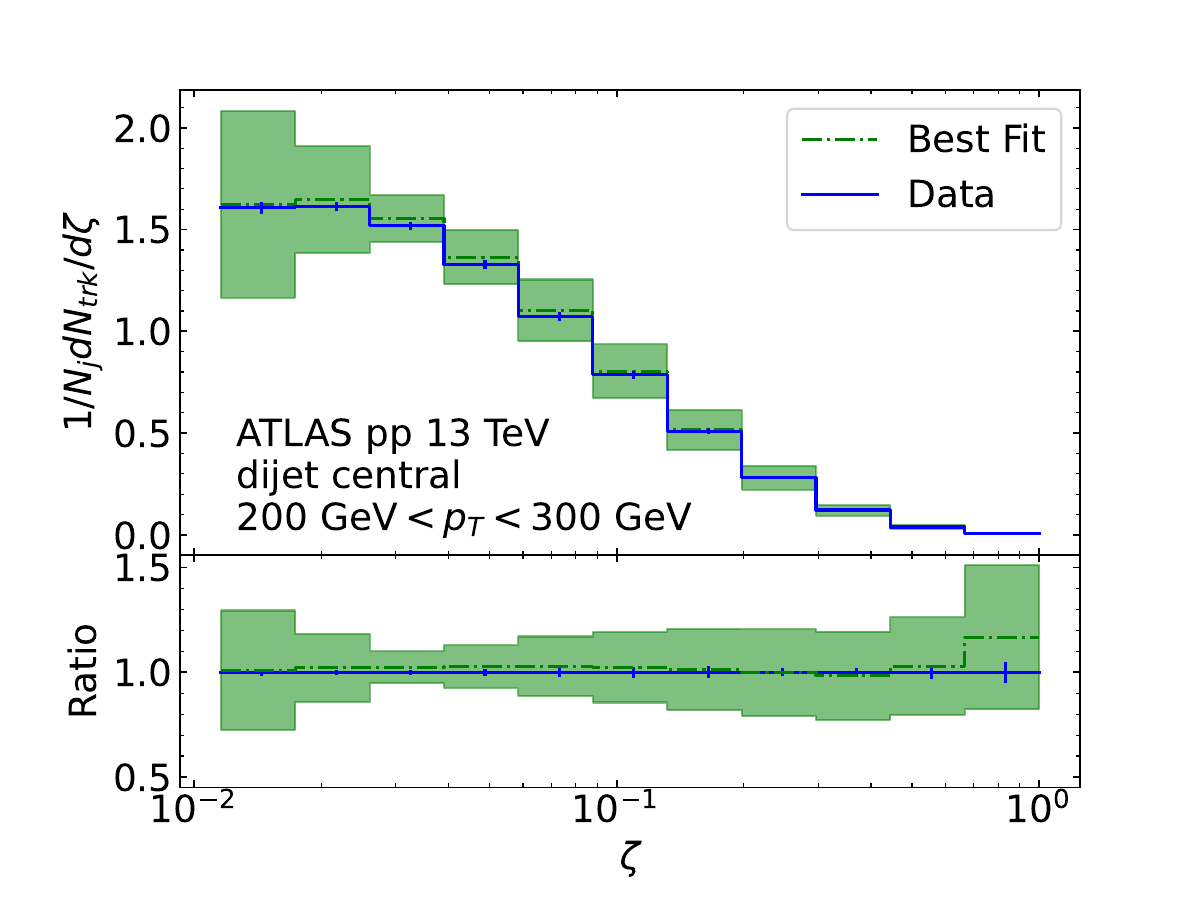}
  \includegraphics[width=0.43\textwidth]{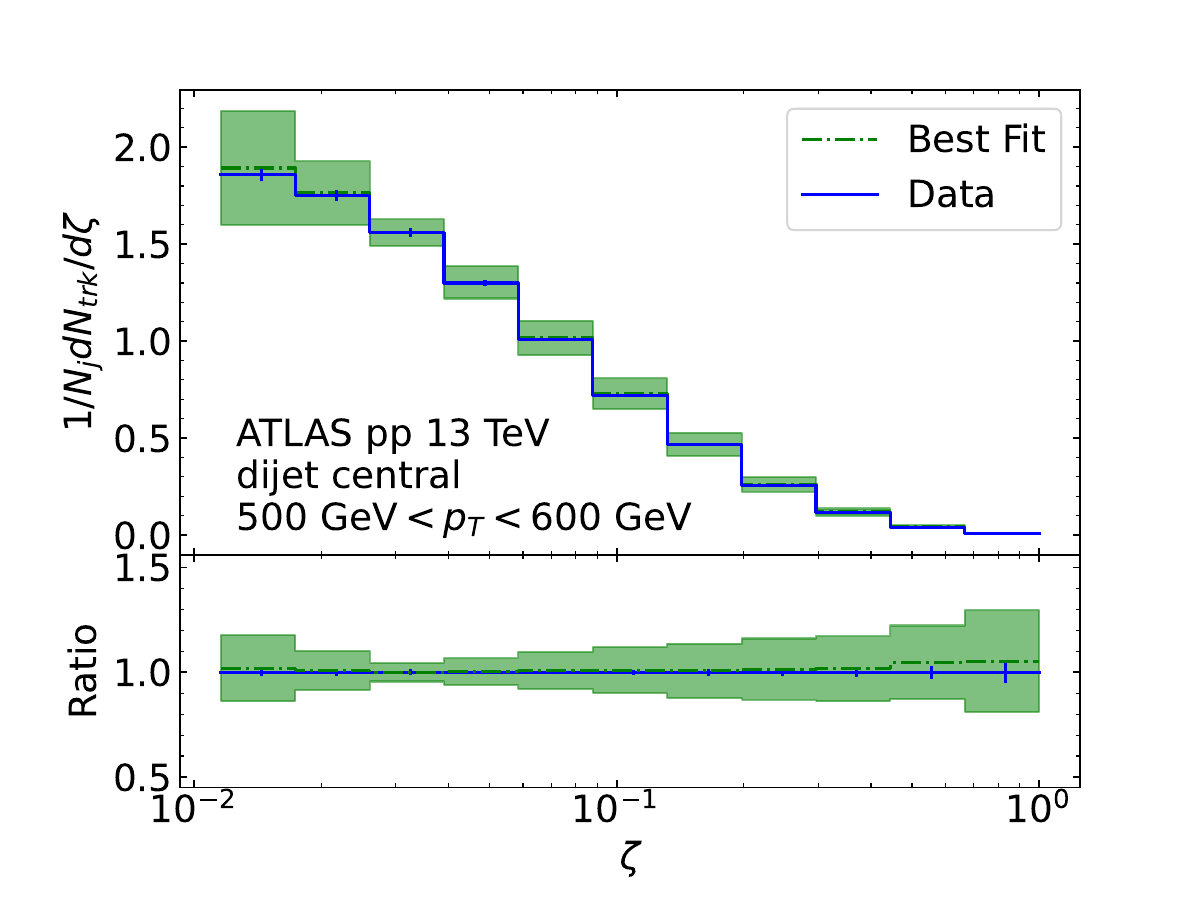}
  \includegraphics[width=0.43\textwidth]{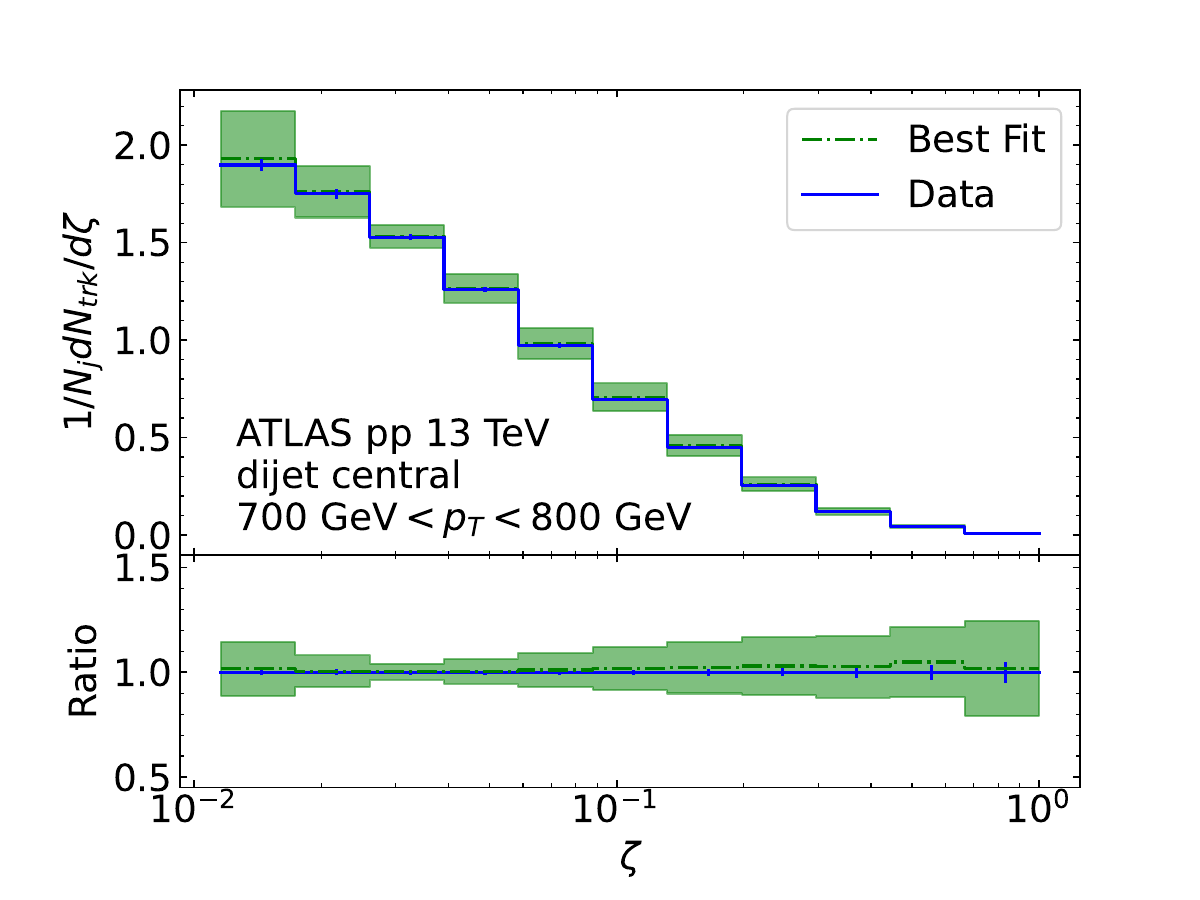}
  \includegraphics[width=0.43\textwidth]{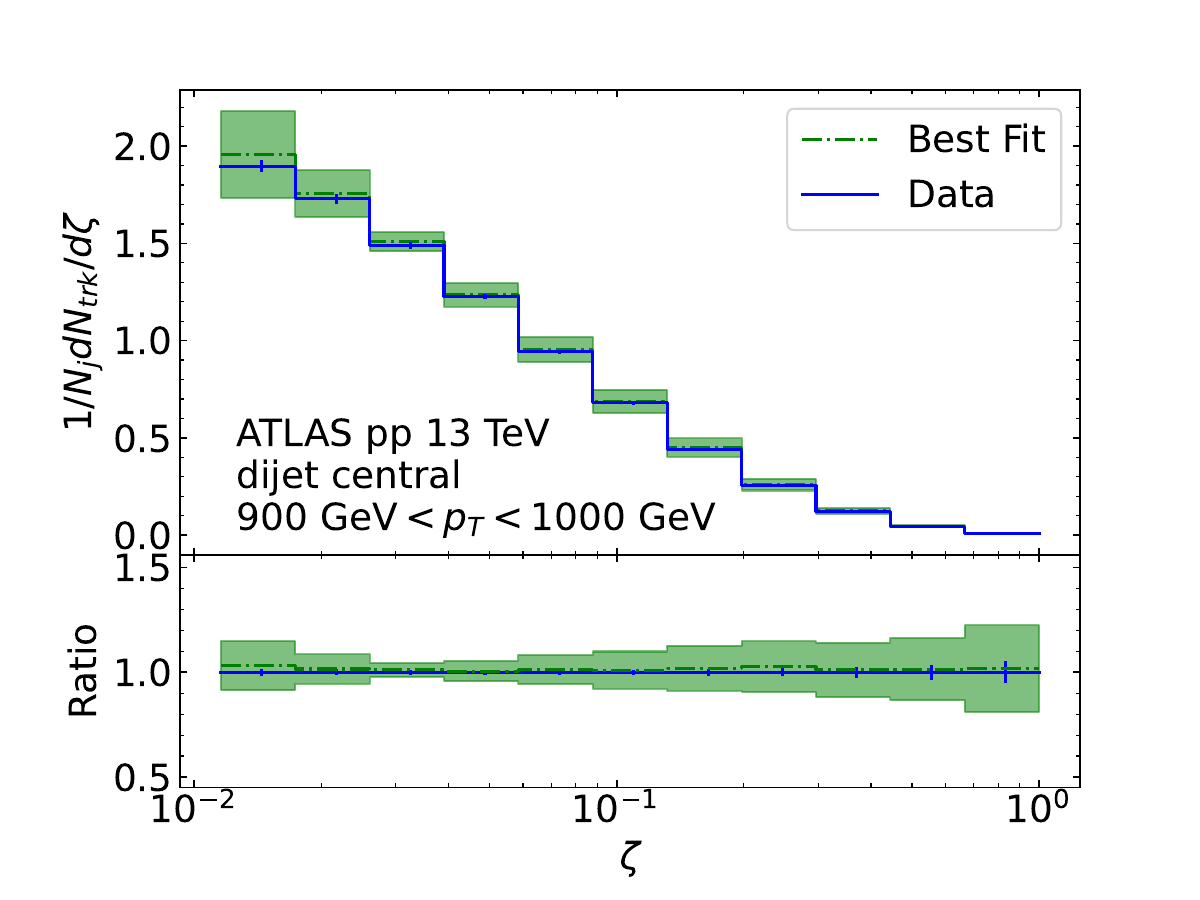}
  \includegraphics[width=0.43\textwidth]{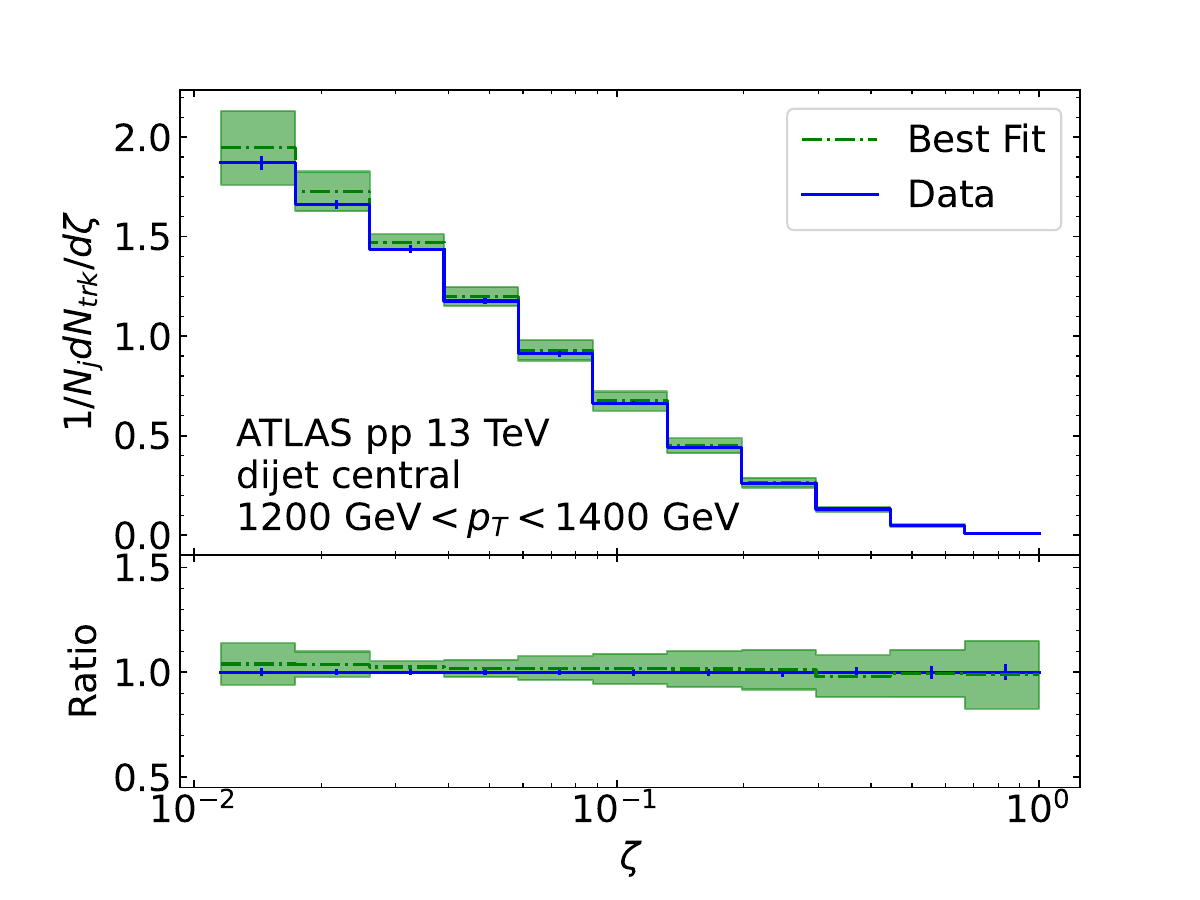}
  \includegraphics[width=0.43\textwidth]{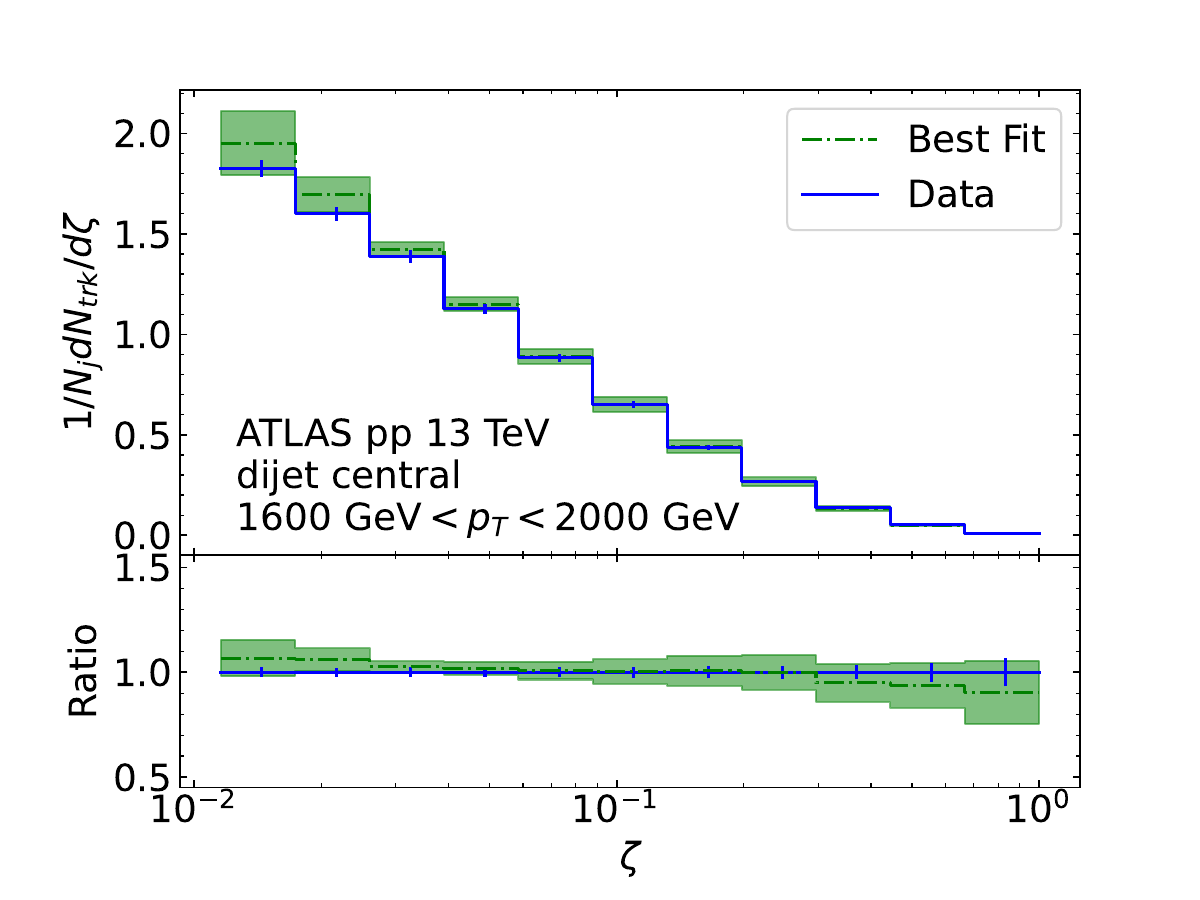}
    \captionsetup{font={stretch=1}}
	\caption{
	Similar to Fig.~\ref{Fig:thdagamma} but for the dijet production and in different bins of transverse momentum of the central jet.
	}
  \label{Fig:thdacjet}
\end{figure}
\begin{figure}[htbp]
  \centering
  \includegraphics[width=0.43\textwidth]{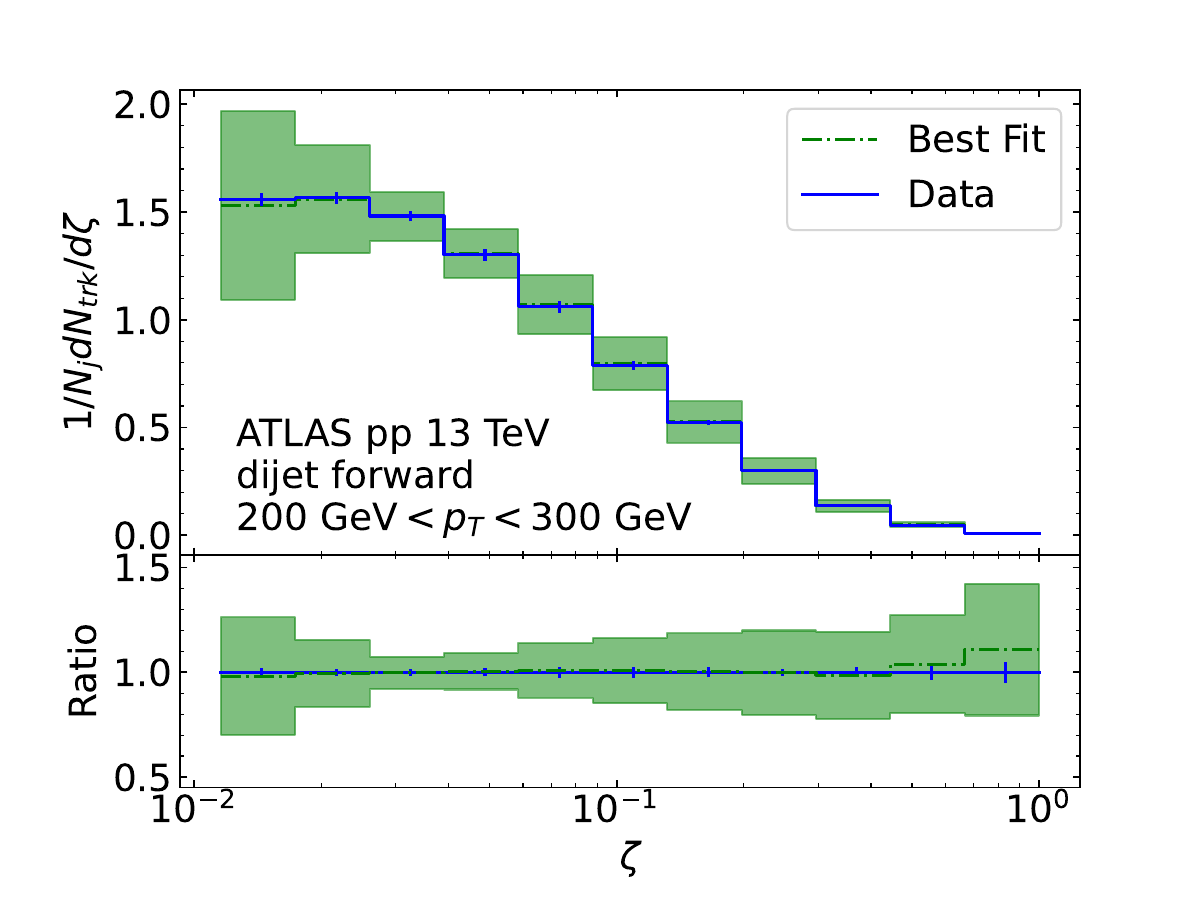}
  \includegraphics[width=0.43\textwidth]{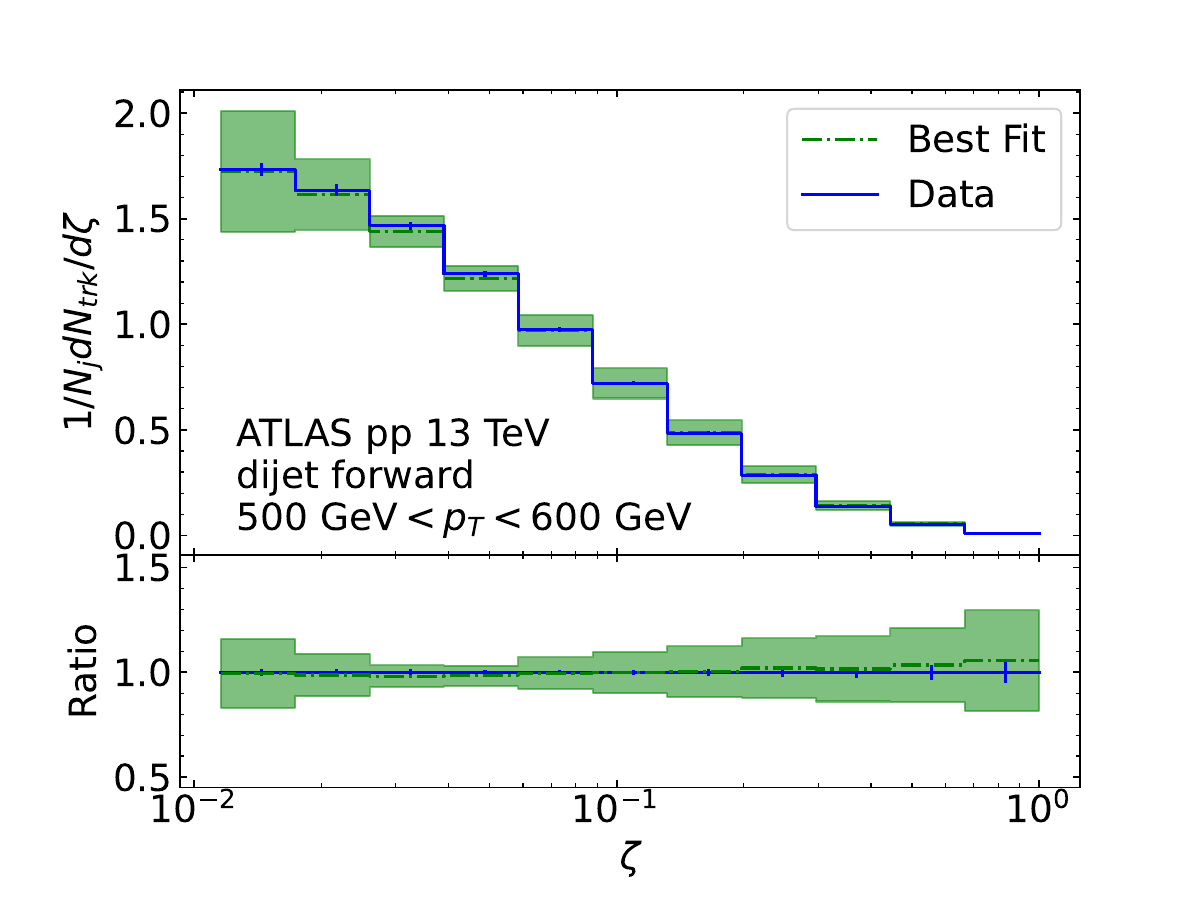}
  \includegraphics[width=0.43\textwidth]{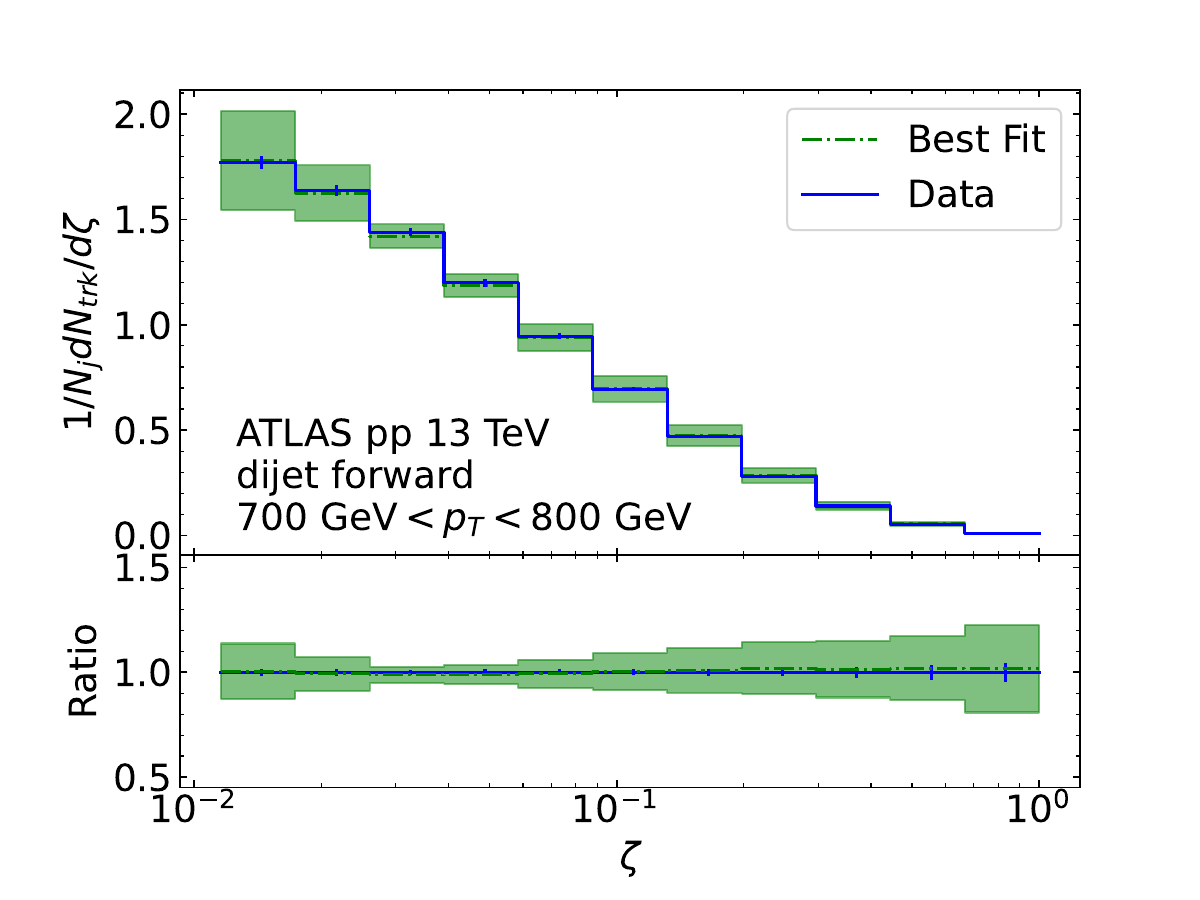}
  \includegraphics[width=0.43\textwidth]{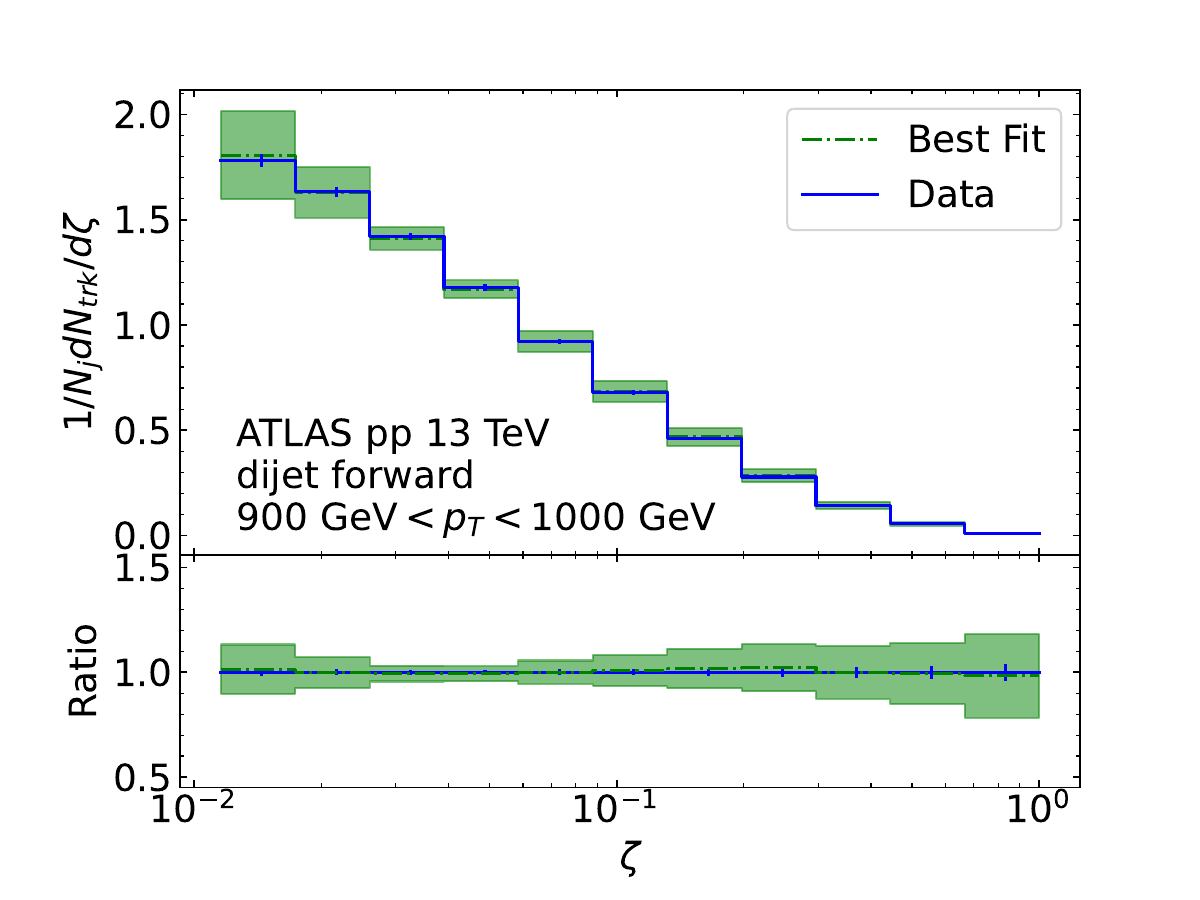}
  \includegraphics[width=0.43\textwidth]{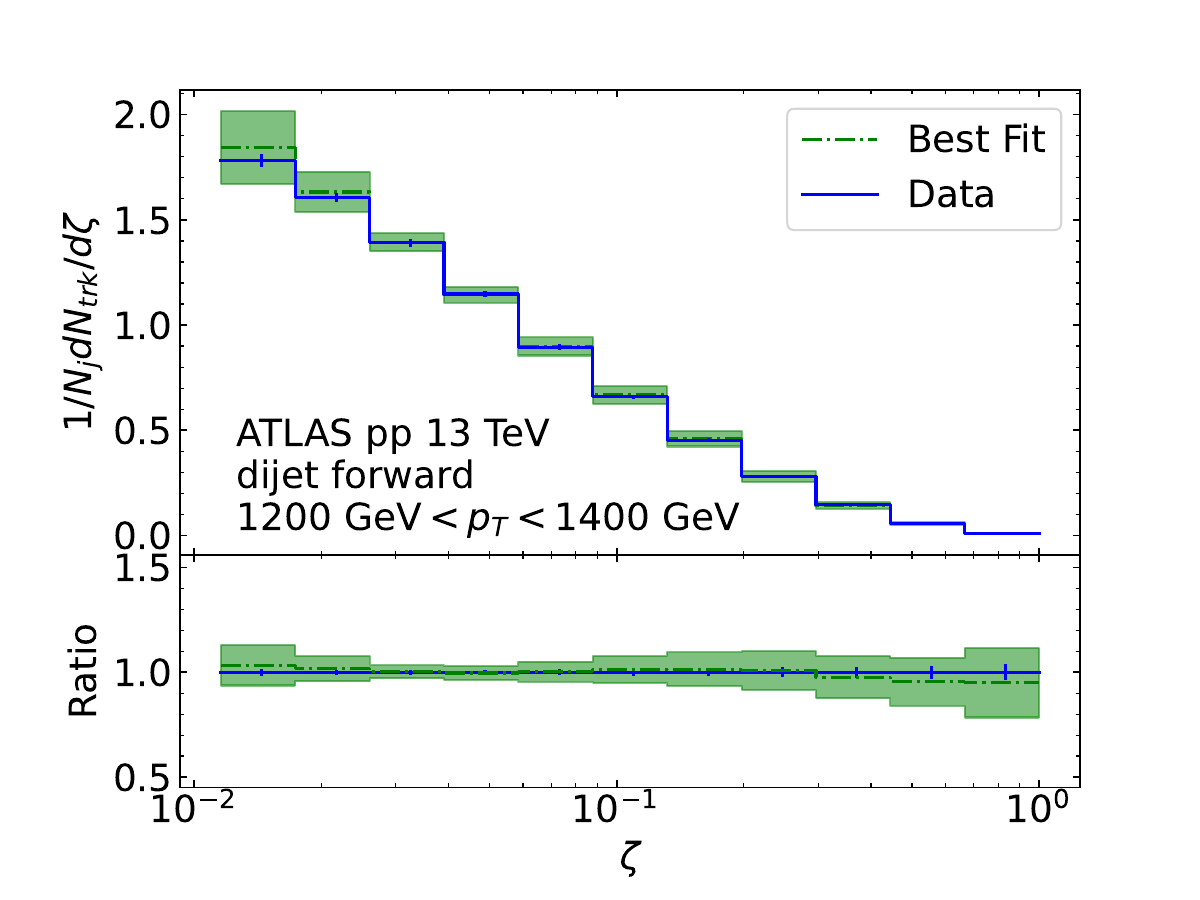}
  \includegraphics[width=0.43\textwidth]{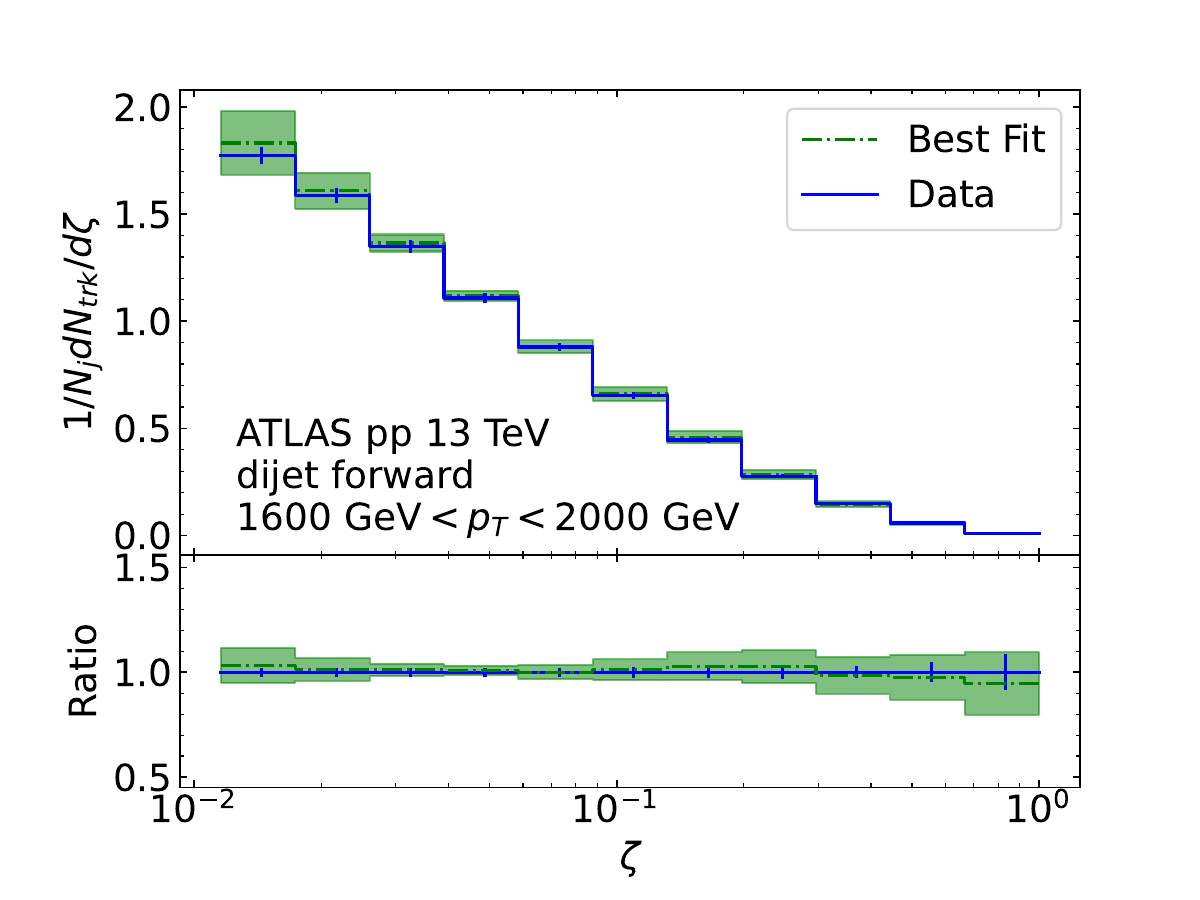}
    \captionsetup{font={stretch=1}}
	\caption{	
	Similar to Fig.~\ref{Fig:thdagamma} but for the dijet production and in different bins of transverse momentum of the forward jet.
	}
  \label{Fig:thdafjet}
\end{figure}
Our NLO predictions based on the best-fit of the fragmentation functions
and their comparison to data are presented in Figs.~\ref{Fig:thdagamma} to~\ref{Fig:thdafjet}. 
In the lower panel of each figure the predictions are normalized to the central value of the experimental data.
The colored bands indicate estimated theoretical uncertainties from scale variations as explained in Sec.~\ref{sec:photon}.
We show the comparison to isolated-photon production in Fig.~\ref{Fig:thdagamma} for both the CMS and ATLAS measurements.
The theoretical predictions locate within 10\% of the data in general with the exception of the first(last) bin in $\xi_T^{\gamma}$($p_{T,h}$) which corresponds to very large $x$ region.   
For the $Z$ boson production shown in Fig.~\ref{Fig:thdazboson}, we again find good agreements to both the CMS and ATLAS measurements. Furthermore, there is consistency observed between the two independent measurements, providing additional confidence in the accuracy of our calculations.
The large $\chi^2$ observed for the ATLAS data is mostly driven by the first $p_{T,h}$ bin of the highest $p_{T,Z}$ region ($>60$ GeV), which has a high precision of a few percents.
Notably it has a $p_{T,h}$ around 4 GeV and receives contributions from region of momentum fractions $x\lesssim 0.01$, and may be affected by additional uncertainties from both theoretical and experimental sides. 
Lastly in Figs.~\ref{Fig:thdacjet}-\ref{Fig:thdafjet} we present comparisons to the ATLAS dijet measurements for both the central and the forward jet and for the selected bins of the transverse momentum of the jet.
We find excellent agreements between our theoretical predictions and the data in the full kinematic region even without considering the theoretical uncertainties.
The scale variations are almost an order of magnitude larger than the experimental uncertainties.

\section{Theory ingredients}
\label{app:Theory ingredients}

We quote the LO unregularized splitting functions below: 
\begin{align}
    P^{(0)}_{qq}(z, \epsilon) = &\ C_F \left[\frac{1 + z^2}{1-z} - \epsilon(1-z) \right]\,,
    \\
    P^{(0)}_{gg}(z, \epsilon) = &\ 2 C_A \frac{(1 - z + z^2)^2}{z (1 - z)}\,,
    \\
    P^{(0)}_{gq}(z, \epsilon) = &\ C_F \left[ \frac{1 + (1-z)^2}{z} - \epsilon z \right]\,,
    \\
    P^{(0)}_{qg}(z, \epsilon) = &\ T_F \left( 1 - \frac{2 z (1 - z)}{1 - \epsilon} \right)\,.
\end{align}
Explicitly, the LO regularized splitting functions are given by
\begin{align}
    P^{+(0)}_{qq}(z) = &\ C_F \left[ \frac{1 + z^2}{[1-z]_+} + \frac{3}{2} \delta(1-z) \right]\,,\label{repqq}
    \\
    P^{+(0)}_{gg}(z) = &\ 2 C_A \left[ \frac{z}{[1-z]_+} + \frac{1-z}{z} + z (1-z) \right] + \delta(1-z) \frac{(11 C_A - 4 n_f T_F)}{6}\,,\label{repgg}
    \\
    P^{+(0)}_{gq}(z) = &\ C_F  \frac{1 + (1-z)^2}{z}\,,\label{repgq}
    \\
    P^{+(0)}_{qg}(z) = &\ T_F \left( z^2+ (1 - z)^2\right)\,.\label{repqg}
\end{align}
For the process of $e^+(p_1)+e^-(p_2)\rightarrow \gamma^*(q)\rightarrow h+X$, the normalized single hadron differential cross sections in $x_h$, where $x_h$ is the energy fraction carried by the tagged hadron (h), can be written as 
\begin{align}
	{1\over \sigma_0}{d\sigma(\gamma^*)\over d x_h} = & {d\hat{\sigma}(\gamma^*)\over d x_g}\otimes D_{h/g}(z,\mu_D)+ {1\over \sum\limits_{j=1}^{n_f}  e^2_{q_j}}\sum_{i=q,\bar q} e^2_{i}\,
	   {d\hat{\sigma}(\gamma^*)\over d x_i}\otimes D_{h/i}(z,\mu_D),
\end{align}
where $\sigma_0=\sum\limits_{i=1}^{n_f} e^2_{q_i}\frac{4\pi\alpha^2 C_A}{3 Q^2}$ with $Q^2=q^2$, and 
\begin{align*}
{d\hat{\sigma}(\gamma^*) \over d x_q}&=\frac{d\hat{\sigma}(\gamma^*)}{d x_{\bar{q}}}= \delta(x_q-1)+{\alpha_S(\mu_R)\over\pi}\bigg[{1\over 2} \,P^{+(0)}_{qq}(x_q)\ln(\frac{Q^2}{\mu_D^2})+ {4 \over 3} \,\left[\frac{\ln(1-x_q)}{1-x_q}\right]_+\\
&-\frac{1}{[1-x_q]_+}+\frac{4\pi^2-27}{9}\,\delta(x_q-1)-\frac{4}{3}\frac{(x_q^2+1)\,\ln(x_q)}{x_q-1}+{1\over3}\,(5-3x_q)\\
&-{2\over3}(x_q+1)\,\ln(1-x_q)\bigg]\,,\\
{d\hat{\sigma}(\gamma^*) \over d x_g}&= {\alpha_S(\mu_R)\over\pi}\, \bigg[P^{+(0)}_{gq}(x_q)\,\ln(\frac{Q^2}{\mu_D^2})+{4 \over 3 }\,(1+(1-x_g)^2)\,\frac{\ln(x_g^2(1-x_g))}{x_g}\bigg]\,.
\end{align*}
Note that $P^{+(0)}_{qq}(x_q)$ and $P^{+(0)}_{gq}(x_g)$ are given by \eqref{repqq} and \eqref{repgq}, respectively. 
% For simplify, here and below we set equal of all quark fragmentation functions.
%

%
For the process of $\mu^+(p_1)+\mu^-(p_2)\rightarrow H(q)\rightarrow h+X$, we have 
\begin{align}
	{1\over \sigma_0}{d\sigma(H)\over d x_h} = & \sum_{i=q,\bar q, g}
	   {d\hat{\sigma}(H)\over d x_i}\otimes D_{h/i}(z,\mu_D),
\end{align}
where
\begin{align*}
\sigma_0&=\frac{\alpha \,\alpha_S(\mu_R)^2 \,C_A C_F \,Q^2 \, C_t^2
   m_{\mu}^2 \left(Q^2-2 m_{\mu }^2\right)}{576 \pi^2 \,s_W^2
   v^2 \,m_W^2 \,\left(m_H^2-Q^2\right)^2}\,,\\
{d\hat{\sigma}(H) \over d x_g}&= 2\delta(x_g-1)+{\alpha_S(\mu_R)\over\pi}\bigg[P^{+(0)}_{gg}(x_g)\,\ln(\frac{Q^2}{\mu_D^2})+2C_A \bigg( \left[\frac{\ln(1-x_g)}{1-x_g}\right]_+\\
&-\frac{23}{36}\,\frac{1}{[1-x_g]_+}
+(\frac{\pi^2}{3}+\frac{349}{108})\,\delta(x_g-1)+\frac{\ln(x_g^2)}{1-x_g}+\frac{23}{36}\,(x_g^2+x_g+1)\\
&+(\frac{1}{x_g}-x_g^2+x_g-2)\,\ln(x_g^2(1-x_g))\bigg)\bigg]\,,\\
{d\hat{\sigma}(H) \over d x_q}&=\frac{d\hat{\sigma}(H)}{d x_{\bar{q}}}= {\alpha_S(\mu_R)\over\pi} \bigg[P^{+(0)}_{qg}(x_q)\,\ln(\frac{Q^2}{\mu_D^2})+{1\over2}(x_q^2+(1-x_q)^2) \,\ln(x_q^2(1-x_q))\\
&+\frac{1}{4}\,x_q(4-7x_q)\bigg]\,.
\end{align*}
Similarly, $P^{+(0)}_{gg}(x_q)$ and $P^{+(0)}_{qg}(x_g)$ are given by \eqref{repgg} and \eqref{repqg}, respectively.

\bibliographystyle{JHEP}
\bibliography{reference}

\end{document}